\documentclass[aps,prd,twocolumn,nofootinbib,showpacs,superscriptaddress]{revtex4-1}

\usepackage{amsfonts}
\usepackage{amsmath}
\usepackage{amssymb}
\usepackage{bm}
\usepackage{dcolumn}
\usepackage{epsfig}
\usepackage{graphicx}
\usepackage{graphics}
\usepackage[latin1]{inputenc}
\usepackage{latexsym}
\usepackage{rotating}
\usepackage[colorlinks=true]{hyperref}
\usepackage[usenames]{color}
\usepackage{yfonts}
\usepackage{xspace} 
\usepackage{mathrsfs}
\usepackage{subfigure}
\usepackage{enumitem}
\usepackage{tabularx}
\usepackage{booktabs}
\usepackage{siunitx}
\usepackage{array}
\usepackage[normalem]{ulem}
\newcolumntype{L}[1]{>{\raggedright\let\newline\\\arraybackslash\hspace{0pt}}m{#1}}
\newcolumntype{C}[1]{>{\centering\let\newline\\\arraybackslash\hspace{0pt}}m{#1}}
\newcolumntype{R}[1]{>{\raggedleft\let\newline\\\arraybackslash\hspace{0pt}}m{#1}}

\usepackage{ulem}
\definecolor {darkgreen}{rgb}{0.2,0.7,0.2}
\graphicspath{{../Figures/}}

\newcommand{\be}{\begin{equation}}
\newcommand{\ee}{\end{equation}}
\newcommand\ba{\begin{eqnarray}}
\newcommand\bse{\begin{subequations}}
\newcommand\ea{\end{eqnarray}}
\newcommand\ese{\end{subequations}}

\newcommand{\order}[1]{\mathcal{O}\left(#1\right)}
\newcommand{\avg}[1]{\left \langle #1 \right \rangle}
\newcommand{\tn}{\tilde{\nabla}}
\newcommand{\tu}{\tilde{u}}

\newcommand{\mat}{{\mbox{\tiny mat}}}

\newcommand{\GR}{{\mbox{\tiny GR}}}
\newcommand{\ST}{{\mbox{\tiny ST}}}
\newcommand{\EA}{{\mbox{\tiny \AE}}}
\newcommand{\VT}{{\mbox{\tiny \AE}}}
\newcommand{\TT}{{\mbox{\tiny TT}}}

\newcommand{\LL}{{\mbox{\tiny LL}}}

\newcommand{\GH}{{\mbox{\tiny GH}}}

\begin{document}
\title{The Gravitational Wave Stress-Energy (pseudo)-Tensor in Modified Gravity}

\author{Alexander Saffer}
\affiliation{eXtreme Gravity Institute, Department of Physics, Montana State University, Bozeman, MT 59717, USA.}

\author{Nicol\'as Yunes}
\affiliation{eXtreme Gravity Institute, Department of Physics, Montana State University, Bozeman, MT 59717, USA.}

\author{Kent Yagi}
\affiliation{Department of Physics, University of Virginia, Charlottesville, Virginia 22904, USA.}
\affiliation{Department of Physics, Princeton University, Princeton, New Jersey 08544, USA.}

\date{\today}

\begin{abstract} 

The recent detections of gravitational waves by the advanced LIGO and Virgo detectors open up new tests of modified gravity theories in the strong-field and dynamical, extreme gravity regime.
Such tests rely sensitively on the phase evolution of the gravitational waves, which is controlled by the energy-momentum carried by such waves out of the system.  
We here study four different methods for finding the gravitational wave stress-energy pseudo-tensor in gravity theories with any combination of scalar, vector, or tensor degrees of freedom.
These methods rely on
the second variation of the action under short-wavelength averaging, 
the second perturbation of the field equations in the short-wavelength approximation, 
the construction of an energy complex leading to a Landau-Lifshitz tensor, and
the use of Noether's theorem in field theories about a flat background. 
We apply these methods in General Relativity, scalar-tensor theories and Einstein-\AE{}ther theory to find the gravitational wave stress-energy pseudo-tensor and calculate the rate at which energy and linear momentum is carried away from the system. 
The stress-energy tensor and the rate of linear momentum loss in Einstein-\AE{}ther theory are presented here for the first time. 
We find that all methods yield the same rate of energy loss, although the stress-energy pseudo-tensor can be functionally different. 
We also find that the Noether method yields a stress-energy tensor that is not symmetric or gauge-invariant, and symmetrization via the Belinfante procedure does not fix these problems because this procedure relies on Lorentz invariance, which is spontaneously broken in Einstein-\AE{}ther theory. 
The methods and results found here will be useful for the calculation of predictions in modified gravity theories that can then be contrasted with observations. 
\end{abstract}

\maketitle
\allowdisplaybreaks[4]

\section{Introduction}
\label{Intro}
With the recent announcements of the discovery of gravitational waves~\cite{Abbott:2016blz,Abbott:2016nmj,TheLIGOScientific:2016pea,Abbott:2017vtc,Abbott:2017oio,TheLIGOScientific:2017qsa}, we are on the brink of a new era in astrophysical science.  We now have evidence that there exist events where two black holes or neutron stars collide and emit, as a consequence of this merger, powerful bursts of gravitational radiation.  The information contained in these gravitational waves provide information about the compact bodies that formed the waves during the inspiral and merger.  In addition, we can learn about how gravitational physics behaves in the \emph{extreme gravity} regime, an area where velocities and gravitational effects are large compared to the surrounding spacetime.  This regime provides an excellent test bed for investigating and constraining the predictions of gravitational theories.

The detection of gravitational waves (GWs) and the science we can extract from them depends sensitively on the models used to extract these waves from the noise.  Because of the way GW detectors work, the extraction relies on an accurate modeling of the rate of change of the GW phase and frequency. In binary systems, this is calculated from the balance law between the rate of change of the binary's binding energy and the energy and momentum extracted by all propagating degrees of freedom. In turn, the latter is obtained from the GW stress-energy (pseudo) tensor (GW SET), which can be calculated in a variety of ways. 

The GW SET was first found in general relativity (GR) by Isaacson in the late 1960's ~\cite{Isaacson:1967zz,Isaacson:1968zza} using what we will call the \emph{perturbed field equations method}.  This method consists of perturbing the Einstein field equations to second order in the metric perturbation about a generic background. The first-order field equations that result describe the evolution of the gravitational radiation. The right-hand side of the second-order field equations yield the GW SET (see Sec.~\ref{sec:GRb} for a detailed explanation).

Since Isaacson's work, other methods for finding the GW SET have been developed.
Stein and Yunes described what we will call the \emph{perturbed action method}, which consists of varying the gravitational action to second order with respect to a generic background~\cite{Stein:2010pn}. Once the variation has been taken, they use short-wavelength averaging to isolate the leading-order contribution to the GW SET.  
Landau and Lifshitz developed a method that consists of constructing a pseudo-tensor from tensor densities with certain symmetries such that its partial divergence vanishes, leading to a conservation law~\cite{Landau:1982dva}. 
The final method we investigate is one that makes use of Noether's theorem, which asserts that the diffeomorphism invariance of a theory automatically leads to the conservation of a tensor~\cite{Noether:1918zz}. This canonical energy-momentum tensor, however, is not guaranteed to be symmetric in its indices or gauge invariant, problems that can be resolved through a symmetrization procedure proposed by Belinfante~\cite{BELINFANTE1940449,Soper:1976bb,Guarrera:2007tu}. 

The four methods described in this paper all give the same GW SET in GR, but this needs not be so in other gravity theories. Scalar-tensor  theories, originally proposed by Jordan~\cite{Jordon}, Fierz~\cite{Fierz:1956zz}, and Brans and Dicke~\cite{Brans:1961sx}, and certain vector-tensor theories, such as Einstein-\AE{}ther theory~\cite{Jacobson:2000xp,Jacobson:2008aj}, are two examples where \emph{a priori} it is not obvious that all methods will yield the same GW SET. In scalar-tensor theories, the GW SET was first computed by Nutku~\cite{Nutku:1969ApJ} using the Landau-Lifshitz method. In Einstein-\AE{}ther theory, the symmetric GW SET has not yet been calculated with any of the methods discussed above; Eling~\cite{Eling:2005zq} derived a non-symmetric GW SET using the Noether current method without applying perturbations. Foster~\cite{Foster:2006az} perturbed the field equations \emph{a la} Isaacson and then used the Noether charge method of~\cite{Wald:1993nt,Iyer:1994ys} to find the rate of change of energy carried by all propagating degrees of freedom.

We here use all four methods discussed above to calculate the GW SET in GR, scalar-tensor and Einstein-\AE{}ther theory. In the GR and scalar-tensor cases, we find that all methods yield exactly the same GW SET. In the Einstein-\AE{}ther case, however, the perturbed field equations method, the perturbed action method and the Landau-Lifshitz method yield slightly different GW SETs; however, the observable quantities computed from them (such as the rate at which energy is removed from the system by all propagating degrees of freedom) are exactly the same. On the other hand, the Noether method applied to Einstein-\AE{}ther theory does not yield a symmetric or gauge invariant GW SET. Symmetrization of the Noether GW SET through the Belinfante procedure fails to fix these problems. This is because this procedure relies on the action being Lorentz invariant at the level of the perturbations, while Einstein-\AE{}ther theory is constructed so as to spontaneously break this symmetry. 

The results obtained here are relevant for a variety of reasons. First, we show for the first time how different methods for the calculation of the GW SET yield the same observables in three gravity theories in almost all cases. Second, we clarify why the Noether method and its Belinfante improvement can fail in Lorentz-violating theories. Third, we provide expressions for the GW SET in Einstein-\AE{}ther theory for the first time, and from this, we compute the rate of energy and linear momentum carried away by all propagating degrees of freedom. Fourth, the methods presented here can be used to compute the rate of change of angular momentum due to the emission of gravitational, vector and scalar waves in Einstein-\AE{}ther theory; in turn, this can be used to place constraints on this theory with eccentric binary pulsar observations. Fifth, the methods described here in detail can be straightforwardly used in other gravity theories that activate scalar and vector propagating modes.

This paper is organized as follows. Section~\ref{sec:GR} focuses on the GW SET in GR, with each subsection~\ref{sec:GRa},~\ref{sec:GRb},~\ref{sec:GRc},~\ref{sec:GRd} detailing a separate method, and the last subsection~\ref{sec:GR_E} calculating physical observables derived from the GW SET. Sections~\ref{sec:ST} and~\ref{sec:EA} apply all the methods for finding the GW SET to scalar-tensor theories and to Einstein-\AE{}ther theory respectively. We choose the common metric signature $(-,+,+,+)$ with units in which $c=1$ and the conventions of~\cite{Carroll:2004st} in which Greek letters in tensor indices stand for spacetime quantities. 

\section{General Relativity}
\label{sec:GR}
In this section, we review how to construct the GW SET in GR and establish notation. We begin by describing the perturbed action method, following mostly~\cite{Stein:2010pn}. We then describe the perturbed field equation method, following mostly~\cite{Carroll:2004st} and the perturbation treatment of~\cite{Isaacson:1967zz,Isaacson:1968zza,Maggiore}. We then discuss the Landau-Lifshitz method, following mostly~\cite{Landau:1982dva,PW} and conclude with a discussion of the Noether method, following mostly~\cite{Maggiore}. 

In all calculations in this section, we work with the Einstein-Hilbert action
\begin{equation}
\label{GR_Action}
S_{\GR}=\dfrac{1}{16 \pi G} \int d^4x \; \sqrt{-g} \; R
\end{equation}
in the absence of matter, where $G$ is Newton's constant, $g$ is the determinant of the four-dimensional spacetime metric $g_{\alpha \beta}$ and $R$ is the trace of the Ricci tensor $R_{\alpha \beta}$.  
By varying the action with respect to the metric, we find the field equations for GR to be
\begin{equation}
\label{eq:GR_FE}
G_{\alpha \beta}=0 \,,
\end{equation}
where $G_{\alpha \beta}$ is the Einstein tensor associated with $g_{\alpha \beta}$.
\subsection{Perturbed Action Method}
\label{sec:GRa}
Let us begin by using the metric decomposition
\begin{equation}
\label{eq:GRa_LinGrav}
g_{\alpha \beta} = \tilde{g}_{\alpha \beta} + \epsilon \; h_{\alpha \beta} + {\cal{O}}(\epsilon^{2}),
\end{equation}
where $\epsilon \ll 1$ is an order counting parameter. One should think of the metric perturbation $h_{\alpha \beta}$ as high frequency ripples on the background $\tilde{g}_{\alpha \beta}$, which varies slowly over spacetime; henceforth, all quantities with an overhead tilde represent background quantities. Both of these fields will be treated as independent, and thus, one could vary the expanded action with respect to each of them separately. For our purposes, however, it will suffice to consider the variation of the expanded action with respect to the background metric order by order in $h_{\alpha \beta}$. 

The action expanded to second order in the metric perturbation is
\begin{equation}
\label{eq:GR_Action-exp}
S_{\GR}=S_{\GR}^{(0)} + \epsilon S_{\GR}^{(1)} + \epsilon^2 S_{\GR}^{(2)} + \order{\epsilon^3}\,,
\end{equation}
where
\begin{subequations}
\label{eq:GRa_ActionExpansion}
\begin{align}
\label{eq:GRa_FinActa}
S^{\tiny (0)}_{\GR} &=\dfrac{1}{16 \pi G} \int d^4x \; \sqrt{-\tilde{g}} \; \tilde{R} \, ,
\\
\label{eq:GRa_FinActb}
S^{\tiny (1)}_{\GR} &=-\dfrac{1}{16 \pi G} \int d^4x \sqrt{-\tilde{g}}
\left[\bar{h}^{\alpha \beta}\tilde{R}_{\alpha \beta} - \frac{1}{2}\tilde{\square}\bar{h}\right. 
\nonumber \\ 
&\left. - \tilde{\nabla}_\alpha \tilde{\nabla}_\beta \bar{h}^{\alpha \beta}\right] \, ,
\\
\label{eq:GRa_FinActc}
S^{\tiny (2)}_{\GR}&=\frac{1}{16 \pi G}\int d^4x \sqrt{-\tilde{g}} \left[\frac{1}{8} \tilde{\nabla}_{\alpha} \bar{h} \tilde{\nabla}^\alpha \bar{h} \right. \nonumber \\
&\left. + \frac{1}{2}\tilde{\nabla}_\beta \bar{h}_{\alpha \gamma} \tilde{\nabla}^\gamma \bar{h}^{\alpha \beta} - \dfrac{1}{4}\tilde{\nabla}_\gamma \bar{h}_{\alpha \beta} \tilde{\nabla}^\gamma \bar{h}^{\alpha \beta}\right]\,,
\end{align}
\end{subequations}
and where we have introduced the trace-reversed metric perturbation
\begin{equation}
\label{eq:GRa_hbar}
\bar{h}^{\alpha \beta}=h^{\alpha \beta}-\frac{1}{2} \tilde{g}^{\alpha \beta}h \,.
\end{equation}
 
Before proceeding, let us simplify the expanded action further by noticing that certain terms do not contribute to observable quantities computed from the GW SET measured at spatial infinity or under short-wavelength averaging. 
Equations~\eqref{eq:GRa_FinActa} and~\eqref{eq:GRa_FinActb} are exact, while we have adjusted~\eqref{eq:GRa_FinActc} for the reasons that follow.  First, there is no term explicitly dependent on the background curvature.  This is because the GW SET will be later used to calculate the rate at which energy and linear momentum are carried away by all propagating degrees of freedom at spatial infinity. These rates will not depend on the background curvature tensor, as the latter vanishes at spatial infinity.  Second, total divergences generated by integration by parts also become boundary terms that will not contribute to observables at spatial infinity.  Third, terms which are odd in the number of perturbation quantities will not contribute to the GW SET after short-wavelength averaging (see App.~\ref{App:Averaging}). From these considerations, we note that $S^{(1)}_\GR$ will not contribute at all to the GW SET, and can therefore be neglected.

We are now left with a simplified effective action 
\begin{equation}
\label{eq:GRa_EffAct}
S_{\GR}^{\tiny \mbox{eff}} = S^{(0) \, \tiny \mbox{eff}}_\GR + \epsilon^2 S^{(2) \, \tiny \mbox{eff}}_\GR + \order{\epsilon^3}\,.
\end{equation}
Variation of Eq.~\eqref{eq:GRa_EffAct} with respect to the background metric will yield the field equations.  The variation of $S^{(0) \tiny \mbox{eff}}_\GR$ gives the Einstein tensor for the background metric, and thus, the variation of $S^{(2) \tiny \mbox{eff}}_\GR$ acts as a source, which we identify with the GW SET. The variation of the second-order piece of the action can be written as  
\begin{equation}
\delta S^{(2) \tiny \mbox{eff}}_\GR = \int d^4x \sqrt{-\tilde{g}} \, t_{\alpha \beta}^\GR \, \delta \tilde{g}^{\alpha \beta}\,,
\end{equation}
with the GW SET
\begin{equation}
\label{eq:GRa_GWSETDef}
\Theta_{\alpha \beta}^\GR = -2 \avg{t_{\alpha \beta}^\GR}\,,
\end{equation}
and the angle-brackets stand for short-wavelength averaging. This averaging is necessary because the background geometry contains many wavelengths of oscillation of the metric perturbation, and the details of the latter only induce higher order corrections to the metric in multiple-scale analysis~\cite{Isaacson:1967zz,Isaacson:1968zza}.

Before we can compute the GW SET, we first need to derive the first-order equations of motion for the field $\bar{h}_{\alpha \beta}$.  This is accomplished by varying Eq.~\eqref{eq:GRa_EffAct} with respect to $\bar{h}^{\alpha \beta}$.  The result is 
\begin{equation}
\label{eq:GRa_EoM}
	\tilde{\square}\bar{h}_{\alpha \beta} - 2 \tilde{\nabla}^\gamma \tilde{\nabla}_{(\alpha} {\bar{h}_{\beta) \gamma}} - \frac{1}{2} \tilde{g}_{\alpha \beta} \, \tilde{\square} \bar{h}=0\,.
\end{equation}
Using the Lorenz gauge condition, 
\begin{equation}
\label{eq:GRa_Lorentz}
\tilde{\nabla}_\alpha \bar{h}^{\alpha \beta}=0\,,
\end{equation}
and taking the trace of Eq.~\eqref{eq:GRa_EoM}, we find
\begin{equation}
\label{eq:GRa_TraceFreeCondition}
\tilde{\square} \bar{h}=0\,.
\end{equation}
If we impose the traceless condition $\bar{h}=0$ on some initial hypersurface, then Eq.~\eqref{eq:GRa_TraceFreeCondition} forces the metric to remain traceless on all subsequent hypersurfaces~\cite{Misner:1974qy}. The combination of the Lorentz gauge condition and the trace free condition yields the transverse-traceless (TT) gauge.  Applying this condition to simplify the field equations in Eq.~\eqref{eq:GRa_EoM} gives
\begin{equation}
\label{eq:GRa_Boxh}
\tilde{\square} \bar{h}^{\TT}_{\alpha \beta}=0\,.
\end{equation}

We are now able to compute the GW SET using Eq.~\eqref{eq:GRa_GWSETDef}.  Performing the necessary calculations yields
\begin{align}
\Theta_{\alpha \beta} &= \frac{1}{32 \pi G} \avg{\tn_\alpha \bar{h}^{\mu \nu} \, \tn_\beta \bar{h}_{\mu \nu} - \frac{1}{2}\tn_\alpha \bar{h} \, \tn_\beta \bar{h} \right. \nonumber \\
& \left. + \tn _\gamma \bar{h} \, \tn^\gamma \bar{h}_{\alpha \beta} + 2 \tn_{[\gamma} \bar{h}_{\delta] \beta} \, \tn^\delta {\bar{h}_\alpha}^\gamma}\,.
\end{align}
We are free to apply the TT gauge at this point, which we make use of.  Since the result of the GW SET is averaged through the brackets, we are also able to integrate by parts and eliminate boundary terms.  These two effects give the familiar result,
\begin{equation}
\label{eq:GRa_FinSET}
\Theta_{\alpha \beta}=\frac{1}{32 \pi G}\avg{\tn_\alpha h_{\mu \nu}^\TT \, \tn_\beta h^{\mu \nu}_\TT}\,.
\end{equation}
\subsection{Perturbed Field Equation Method}
\label{sec:GRb}

Let us begin by decomposing the metric $g_{\alpha \beta}$ as
\begin{equation}
\label{eq:GRb_gExp}
g_{\alpha \beta} = \eta_{\alpha \beta} + h^{\tiny (1)}_{\alpha \beta} + h^{\tiny (2)}_{\alpha \beta}\,,
\end{equation}
where $\eta_{\alpha \beta}$ is the Minkowski metric and $(h^{\tiny (1)}_{\alpha \beta},h^{\tiny (2)}_{\alpha \beta})$ are first and second order metric perturbations respectively. Unlike in the previous section, the background is here not arbitrary but chosen to be the Minkowski spacetime. We assume that $h_{\alpha \beta}^{(n)} = \order{\epsilon^{n}}$, but we no longer keep track explicitly of the order counting parameter $\epsilon$. 

With this decomposition, we now expand the Einstein field equations [Eq.~\eqref{eq:GR_FE}] to second order in the metric perturbation. We obtain the field equations for $h^{\tiny (1)}_{\alpha \beta}$ by truncating the expansion of the field equations to linear order in the perturbation: 
\begin{equation}
\label{eq:GRb_FE-order1}
G_{\alpha \beta}[h^{\tiny (1)}] = 0\,.
\end{equation}
We obtain the GW SET by truncating the expansion to second order in the perturbation:
\begin{equation}
\label{eq:GRb_FE-order2}
G_{\alpha \beta}\left[h^{\tiny (2)}\right] = 8 \pi \Theta_{\alpha \beta} \equiv - \left<G_{\alpha \beta}\left[\left(h^{\tiny (1)}\right)^{2}\right]\right> \,,
\end{equation}
where again the angled-brackets stand for short-wavelength averaging\footnote{Strictly speaking, this method is not identical to Isaacson's original work~\cite{Isaacson:1967zz,Isaacson:1968zza} because here we assume a priori that the background is Minkowski.}.

We begin by expanding the field equations to first order in the metric perturbation. The Ricci tensor takes the form
\begin{equation}
\label{eq:GRb_Ricci1o}
R^{\tiny (1)}_{\alpha \beta}=\frac{1}{2}\left[2\partial_\gamma \partial_{(\alpha}{h_{\beta)}^{(1)\gamma}} - \partial_\alpha \partial_\beta h^{(1)} - \partial_\gamma \partial^\gamma h^{(1)}_{\alpha \beta}\right]\,,
\end{equation}
and plugging this and its trace into the Einstein tensor of Eq.~\eqref{eq:GRb_FE-order1} gives 
\begin{multline}
\label{eq:GRb_FE_Exp1o}
\partial_\gamma \partial_{(\alpha}{h^{\tiny (1)}_{\beta)}}^\gamma - \frac{1}{2}\partial_\alpha \partial_\beta h^{\tiny (1)} - \frac{1}{2}\partial_\gamma \partial^\gamma h^{\tiny (1)}_{\alpha \beta} \\
+\frac{1}{2}\eta_{\alpha \beta}\left(\partial_\gamma \partial^\gamma h^{\tiny (1)} - \partial^\gamma \partial^\delta h^{\tiny (1)}_{\gamma \delta}\right)=0 \,.
\end{multline}
As in Sec.~\ref{sec:GRa}, imposing the Lorenz gauge $\partial^{\alpha} \bar{h}_{\alpha \beta}^{\tiny (1)} = 0$ on the trace-reversed metric perturbation $\bar{h}_{\alpha \beta}^{\tiny (1)} \equiv h_{\alpha \beta}^{\tiny (1)} - (1/2) \, \eta_{\alpha \beta} \, h^{\tiny (1)}$, and taking the Minkowski trace in Eq.~\eqref{eq:GRb_FE_Exp1o} forces $h^{\tiny (1)} \equiv \eta^{\alpha \beta} h_{\alpha \beta}^{\tiny (1)}$ to satisfy a wave equation in flat spacetime. This, in turn, allows us to refine the Lorenz gauge into the TT gauge in flat spacetime to further simplify the field equations to 
\begin{equation}
\label{eq:GRb_TTEoM}
\partial_\gamma \partial^\gamma h^{\tiny (1)\TT}_{\alpha \beta}=0\,.
\end{equation}

The next step is to investigate the field equations to $\order{h^2}$.  The expansion of the Ricci tensor is
\begin{widetext}
	\begin{align}
	\label{eq:GRb_Ricci2o}
	R^{(2)}_{\alpha \beta}&=\frac{1}{2}\left[2\partial_\gamma \partial_{(\alpha}{h_{\beta)}^{(2)\gamma}} - \partial_\alpha \partial_\beta h^{(2)} - \partial_\gamma \partial^\gamma h^{(2)}_{\alpha \beta}\right] + \frac{1}{2} \partial_\beta \left[h^{(1)\gamma \delta}\left(2\partial_{(\alpha}h^{(1)}_{\gamma)\delta}-\partial_\delta h^{(1)}_{\alpha \gamma}\right)\right] -\frac{1}{2}\partial_\gamma \left[h^{(1)\gamma \delta}\left(2\partial_{(\alpha}h^{(1)}_{\beta)\delta}-\partial_\delta h^{(1)}_{\alpha \beta}\right)\right] \nonumber\\
	&-\frac{1}{4}\left[\partial_\alpha h^{(1)\gamma \delta}\partial_\beta h^{(1)}_{\gamma \delta}+2 \partial_\gamma h^{(1)\delta}_{\alpha} \partial_\delta h^{(1)\gamma}_{\beta} - 2\partial_\gamma h^{(1)\delta}_{\alpha} \partial^\gamma h^{(1)}_{\beta \delta}\right] + \frac{1}{4}\partial_\gamma h^{(1)} \left[2\partial_{(\alpha}h^{(1)\gamma}_{\beta)} - \partial^\gamma h^{(1)}_{\alpha \beta}\right]\,.
	\end{align}
\end{widetext}
This equation can be simplified by imposing the Lorenz gauge on $h^{\tiny (2)}_{\alpha \beta}$, namely $\partial^{\alpha} \bar{h}_{\alpha \beta}^{\tiny (2)} = 0$ with $\bar{h}_{\alpha \beta}^{\tiny (2)} \equiv h_{\alpha \beta}^{\tiny (2)} - (1/2) \, \eta_{\alpha \beta} \, h^{\tiny (2)}$ and $h^{\tiny (2)} \equiv \eta^{\alpha \beta} h_{\alpha \beta}^{\tiny (2)}$. As before, we can refine this gauge into the TT gauge by setting $h^{\tiny (2)}=0$, which is compatible with $R^{(2)} = 0$. With this at hand, the Ricci tensor in the TT gauge reduces to 
\begin{align}
\label{eq:GRb_Ricci2}
R^{(2)}_{\alpha \beta} &= -\frac{1}{2} \partial_\gamma \partial^\gamma h^{(2)\TT}_{\alpha \beta}
+ \frac{1}{4} \partial_\alpha h^{\tiny (1) \gamma \delta}_{\TT} \partial_\beta h^{\tiny (1)\TT}_{\gamma \delta} \nonumber \\
&+ \frac{1}{2}h^{\tiny (1)\gamma \delta}_\TT \partial_\alpha \partial_\beta h^{\tiny (1)\TT}_{\gamma \delta} + \frac{1}{2}h^{\tiny (1)\gamma \delta}_\TT \partial_\gamma \partial_\delta h^{\tiny (1)\TT}_{\alpha \beta} \nonumber \\
&- h^{\tiny (1)\gamma \delta}_\TT \partial_\delta \partial_{(\alpha}h^{\tiny (1)\TT}_{\beta) \gamma}+ \partial_{[\delta}h_{\gamma]\beta}^{\tiny (1)\TT} \partial^\delta h_{\alpha}^{\tiny (1)\TT \gamma}\,.
\end{align}
The first line of Eq.~\eqref{eq:GRb_Ricci2o} (and the first term of Eq.~\eqref{eq:GRb_Ricci2}) is nothing but Eq.~\eqref{eq:GRb_Ricci1o} with $h^{(1)}_{\alpha \beta} \to h^{(2)}_{\alpha \beta}$, which will contribute to the left-hand side of Eq.~\eqref{eq:GRb_FE-order2}. Using Eq.~\eqref{eq:GRb_FE-order2}, the GW SET is then (suppressing superscripts for neatness)
\begin{equation}
\Theta_{\alpha \beta} = \frac{1}{32 \pi G} \avg{\partial_\alpha h^{TT}_{\gamma \delta} \; \partial_\beta h^{\gamma \delta}_{TT}}\,,
\end{equation}
after integrating by parts inside of the averaging scheme. This expression is identical to Eq.~\eqref{eq:GRa_FinSET}, albeit around a Minkowski background.

\subsection{Landau-Lifshitz Method}
\label{sec:GRc}
The Landau-Lifshitz method makes use of a formulation of GR in terms of the ``gothic g metric," a tensor density defined as 
\begin{equation} 
\label{GRc_gothicg}
\textgoth{g}^{\alpha \beta} \equiv \sqrt{-g} \; g^{\alpha \beta} \,.
\end{equation}
With this density, one can construct the 4-tensor density
\begin{equation}
\label{eq:GRc_HTensor}
\mathcal{H}^{\alpha \mu \beta \nu} \equiv \textgoth{g}^{\alpha \beta}\textgoth{g}^{\mu \nu} - \textgoth{g}^{\alpha \nu}\textgoth{g}^{\beta \mu} \,,
\end{equation}
which has the remarkable property that
\begin{equation}
\label{eq:GRc_HProp1}
\partial_\mu \partial_\nu \mathcal{H}^{\alpha \mu \beta \nu} = 2 \left(-g\right)G^{\alpha \beta}+ 16 \pi G \left(-g\right)t^{\alpha \beta}_{\LL}\,,
\end{equation}
where $G^{\alpha \beta}$ is exactly the Einstein tensor and $t_{\LL}^{\alpha \beta}$ is known as the Landau-Lifshitz pseudotensor~\cite{Landau:1982dva},
\begin{align}
\label{eq:GRc_tLL}
(-g)t^{\alpha \beta}_{\LL} &= \frac{1}{16 \pi G}\left[\partial_\gamma \textgoth{g}^{\alpha \beta}\partial_\delta \textgoth{g}^{\gamma \delta}-\partial_\gamma \textgoth{g}^{\alpha \gamma}\partial_\delta \textgoth{g}^{\beta \delta} \right. \nonumber \\
&\left. +\frac{1}{2}g^{\alpha \beta}g_{\gamma \delta}\partial_\epsilon \textgoth{g}^{\gamma \kappa} \partial_\kappa \textgoth{g}^{\delta \epsilon} - 2 g_{\delta \epsilon}g^{\gamma (\alpha}\partial_\kappa \textgoth{g}^{\beta) \epsilon}\partial_\gamma \textgoth{g}^{\delta \kappa} \right. \nonumber \\
&\left. + g_{\gamma \delta}g^{\epsilon \kappa}\partial_\epsilon \textgoth{g}^{\alpha \gamma}\partial_\kappa \textgoth{g}^{\beta \delta} + \frac{1}{8}\left(2g^{\alpha \gamma}g^{\beta \delta}-g^{\alpha \beta}g^{\gamma \delta}\right) \right. \nonumber \\
&\left. \times \left(2g_{\epsilon \kappa}g_{\lambda \sigma}-g_{\kappa \lambda}g_{\epsilon \sigma}\right)\partial_\gamma \textgoth{g}^{\epsilon \sigma}\partial_\delta \textgoth{g}^{\kappa \lambda}\right] \,.
\end{align}
 Substituting the GR field equations into Eq.~\eqref{eq:GRc_HProp1} gives
\begin{equation}
\label{eq:GRc_HProp2}
\partial_\mu \partial_\nu \mathcal{H}^{\alpha \mu \beta \nu} = 16 \pi G \left(-g\right) \left(T^{\alpha \beta}_{\mat} + t^{\alpha \beta}_{\LL}\right)\,,
\end{equation}
where $T^{\alpha \beta}_{\mat}$ is the matter SET. We have kept this term here for clarity, but in this paper $T^{\alpha \beta}_{\mat} = 0$. 

We can now use the symmetries of $\mathcal{H}^{\alpha \mu \beta \nu}$ to derive some conservation laws. First, we notice that by construction $\mathcal{H}^{\alpha \mu \beta \nu}$ has the same symmetries as the Riemann tensor. 
Since $\mathcal{H}^{\alpha \mu \beta \nu}$ is antisymmetric in $\alpha$ and $\mu$, we find 
\begin{equation}
\label{eq:GRc_Hconserv}
\partial_\alpha \partial_\mu \partial_\nu \mathcal{H}^{\alpha \mu \beta \nu}=0 \,.
\end{equation}
Equation~\eqref{eq:GRc_Hconserv} implies there exists a conserved quantity, which we will define as
\begin{equation}
\label{eq:GRc_pseudoSET}
	T^{\alpha \beta} \equiv \frac{1}{16 \pi G} \left(\partial_\mu \partial_\nu \mathcal{H}^{\alpha \mu \beta \nu}\right) \,.
\end{equation}
When this quantity is short-wavelength averaged, one recovers the GW SET
\begin{equation}
\label{eq:GRc_ThetaEqn}
\Theta^{\alpha \beta}=\avg{T^{\alpha \beta}} = \avg{\left(-g\right) t^{\alpha \beta}_{\LL}}\,.
\end{equation}

What we described above is fairly general, so now we evaluate the GW SET in terms of a metric perturbation from a Minkowski background. Using the expansion
\begin{equation}
g_{\alpha _\beta} = \eta_{\alpha \beta} + h_{\alpha \beta}\,,
\end{equation}
with $h_{\alpha \beta} \sim \order{\epsilon}$, the gothic metric becomes
\begin{equation}
\label{eq:GRc_gothg2}
\textgoth{g}^{\alpha \beta}=\eta^{\alpha \beta}-\bar{h}^{\alpha \beta} + \mathcal{O}(h^2)\,,
\end{equation}
where $\bar{h}^{\alpha \beta}$ is the trace reversed metric perturbation.  Equation~\eqref{eq:GRc_tLL} can now be simplified and written to second order in the metric perturbation as
\begin{align}
\label{eq:GRc_tLL2}
\left(-g\right)t^{\alpha \beta}_{\LL} &= \frac{1}{16 \pi G}\left[\partial_\gamma \bar{h}^{\alpha \beta}\partial_\delta \bar{h}^{\gamma \delta}-\partial_\gamma \bar{h}^{\alpha \gamma}\partial_\delta \bar{h}^{\beta \delta} \right. \nonumber \\
&\left. + \frac{1}{2}\eta^{\alpha \beta}\partial_\gamma {\bar{h}^\delta}_\epsilon \partial_\delta \bar{h}^{\epsilon \gamma}-2 \partial_\gamma \bar{h}^{\epsilon(\alpha}\partial^{\beta)}{\bar{h}^{\gamma}}_{\epsilon}+\partial^\gamma {\bar{h}^\alpha}_\delta \partial_\gamma \bar{h}^{\beta \delta} \right. \nonumber \\
&\left. + \frac{1}{2}\partial^{\alpha}\bar{h}_{\gamma \delta} \partial^\beta \bar{h}^{\gamma \delta} - \frac{1}{4}\partial^\alpha \bar{h} \partial^\beta \bar{h}-\frac{1}{4}\eta^{\alpha \beta}\partial^\gamma \bar{h}_{\delta \epsilon} \partial_{\gamma}\bar{h}^{\delta \epsilon} \right. \nonumber\\
&\left. + \frac{1}{8}\eta^{\alpha \beta}\partial^{\gamma}\bar{h}\partial_\gamma \bar{h}\right]\,.
\end{align}
Inserting this expression into Eq.~\eqref{eq:GRc_pseudoSET} we find $T_{\alpha \beta}$, and after short-wavelength averaging, exploiting the gauge freedom to use the TT gauge, integrating by parts, and using the first-order field equations, we obtain
\begin{equation}
\Theta^{\alpha \beta}= \frac{1}{32 \pi G}\left \langle \partial^\alpha h_{\gamma \delta}^{\TT} \partial^\beta h^{\gamma \delta}_{\TT}\right \rangle\,.
\end{equation}
Notice that this method does not allow us to find the first-order field equations, which are typically obtained by the perturbed field equations method of the previous subsection to first order in the metric perturbation.  The final result matches those found in Secs.~\ref{sec:GRa} and~\ref{sec:GRb}.

\subsection{Noether Current Method}
\label{sec:GRd}

Noether showed that symmetries of the action lead to conserved quantities~\cite{Noether:1918zz}, which have become known as Noether currents $j^\alpha$.  Taking a field theoretical approach (for fields propagating on a Minkowski background), consider the action 
\begin{equation}
\label{eq:GRd_Act}
S=\int d^{4}x \; \mathcal{L}(\phi_L,\partial_\alpha \phi_{L})\,,
\end{equation}
where $\phi_L=(\phi_{1},\phi_{2},\ldots,\phi_{L})$ are the fields of the spacetime and ${\cal{L}}$ is a Lagrangian density. From Hamilton's principle, the variation of the action in Eq.~\eqref{eq:GRd_Act} must vanish.  This leads to
\begin{equation}
\label{eq:GRd_VarL2}
\left(\frac{\partial \mathcal{L}}{\partial \phi_L}-\partial_\alpha \frac{\partial \mathcal{L}}{\partial(\partial_\alpha \phi_L)}\right) \delta \phi_L + \partial_\alpha \left(\frac{\partial \mathcal{L}}{\partial(\partial_\alpha \phi_L)} \delta \phi_L\right)=0\,.
\end{equation}
We choose our fields to remain constant on the boundary, which leads to the second term in Eq.~\eqref{eq:GRd_VarL2} to vanish.  This is because the variation still takes place within an integral and total derivatives become boundary terms.  The term remaining constitutes the Euler-Lagrange equations.  These must be satisfied, and give the equations of motion for the fields.

To find the conserved energy-momentum tensor, we vary the Lagrangian density with respect to the coordinates,
\begin{equation}
\frac{\partial \mathcal{L}}{\partial x^\alpha} = \frac{\partial \mathcal{L}}{\partial \phi_L}\partial_\alpha \phi_L + \frac{\partial \mathcal{L}}{\partial \left(\partial_\beta \phi_L \right)} \partial_\beta \partial_\alpha \phi_L \,.
\end{equation}
The first term on the right-hand side can be replaced with the Euler-Lagrange equations.  The result, after basic analysis, is
\begin{align}
\frac{\partial \mathcal{L}}{\partial x^\alpha} - \frac{\partial}{\partial x^\beta}\left(\frac{\partial \mathcal{L}}{\partial \left(\partial_\beta \phi_L\right)} \partial_\alpha \phi_L\right) &= 0 \,,
\end{align}
which then leads to the conservation law $\partial_\alpha j^\alpha{}_\beta=0$ with the current
\begin{equation}
\label{eq:GRd_SET}
{j^\alpha}_\beta = \mathcal{L} \; {\delta^\alpha}_\beta - \frac{\partial \mathcal{L}}{\partial \left(\partial_\alpha \phi_L\right)}\partial_\beta \phi_L\,,
\end{equation}
which is known as the \emph{canonical} SET. The GW SET is typically assumed to be its short-wavelength average:
\begin{equation}
\label{eq:GW_SET-from-can}
\Theta^{\alpha}{}_{\beta} = \left<j^{\alpha}{}_{\beta} \right>\,.
\end{equation}
Unlike the previous methods, which made use of the symmetries of the variations and expansions, there is no mandate that the SET derived here is symmetric. In fact, even in Maxwell's electrodynamics, the canonical SET is not symmetric (see Appendix~\ref{App:Belinfante}).

\if{0}
Emmy Noether showed that symmetries of the action lead to conserved quantities~\cite{Noether:1918zz}, which have become known as Noether currents $j^A$.  The exponent $A$ represents the number of indices ($A=\{\alpha, \alpha \beta, \alpha \beta \gamma, ... \}$) that the current has.  Taking a field theoretical approach about a fixed background, consider the action
\begin{equation}
\label{eq:GRd_Act}
A=\int dx \; \mathcal{L}(\phi_L,\partial_\alpha \phi_{L},\partial_\alpha \partial_\beta \phi_L, ...)\,,
\end{equation}
where $\phi_L$ represents the fields of the spacetime.  From Hamilton's principle, it is known that the variation of the Lagrangian must vanish in Eq.~\eqref{eq:GRd_Act}.  We may then construct a new term $\Upsilon^A$ that satisfies the relation 
\begin{equation}
\label{eq:GRd_VarLB}
\delta \mathcal{L}=\partial_A \delta \Upsilon^A = 0 \,.
\end{equation}
In addition to this, Hamilton's principle can also give
\begin{equation}
\label{eq:GRd_VarL1}
\delta \mathcal{L} = \frac{\partial \mathcal{L}}{\partial \phi_L} \delta \phi_L+ \frac{\partial \mathcal{L}}{\partial(\partial_A \phi_L)} \delta (\partial_A \phi_L)\,.
\end{equation}
Taking Eq.~\eqref{eq:GRd_VarL1} and integrating by parts yields 
\begin{equation}
\label{eq:GRd_VarL2}
\delta \mathcal{L}= \left(\frac{\partial \mathcal{L}}{\partial \phi_L}-\partial_A \frac{\partial \mathcal{L}}{\partial(\partial_A \phi_L)}\right) \delta \phi_L + \partial_A \left(\frac{\partial \mathcal{L}}{\partial(\partial_A \phi_L)} \delta \phi_L\right)\,.
\end{equation}
The first term in Eq.~\eqref{eq:GRd_VarL2} is the Euler-Lagrange equations.  If these are satisfied, only the total derivative is left.  Equation~\eqref{eq:GRd_VarLB} and \eqref{eq:GRd_VarL2} can be combined to define the Noether current,
\begin{equation}
\label{eq:GRd_current}
j^A \equiv \delta \Upsilon^A - \frac{\partial \mathcal{L}}{\partial \left(\partial_A \phi_L\right)}\delta \phi_L \,.
\end{equation}
Notice that due to the variations taken, this current must be conserved, and thus satisfies
\begin{equation}
\partial_A j^A=0.
\end{equation}
Eq.~\eqref{eq:GRd_current} is the Noether current that exists for all systems that satisfy the Euler-Lagrange equations found in the first term of Eq.~\eqref{eq:GRd_VarL2}. 

Since the determination of conserved quantities involves the symmetries associated with the action, it is useful to investigate how a simple transformation affects these currents.  Let the transformation from old to new coordinates be given by
\begin{subequations}
	\begin{align}
	x^\alpha &= \tilde{x}^\alpha + X^\alpha \left(\tilde{x},\epsilon\right), \\
	\phi_L &= \tilde{\phi}_L.
	\end{align}
\end{subequations}
Here, $X^\alpha \left(\tilde{x},\epsilon\right)$ is the coordinate transformation function with $\epsilon$ defined as the scale of the transformation.  Under this transformation, $\delta \Upsilon^A$ can be rewritten as
\begin{equation}
\delta \Upsilon^A = \mathcal{L} \frac{\partial X^A}{\partial \epsilon}\delta \epsilon.
\end{equation}
Assuming an infinitesimally small coordinate transformation $X^\alpha = {\delta^\alpha}_\beta \epsilon^\beta$, Eq.~\eqref{eq:GRd_current} becomes
\begin{equation}
\label{eq:GRd_SET}
{j^\alpha}_\beta = \mathcal{L} {\delta^\alpha}_\beta - \frac{\partial \mathcal{L}}{\partial \left(\partial_\alpha \phi_L\right)}\partial_\beta \phi_L\,.
\end{equation}
Eq.~\eqref{eq:GRd_SET} is known as the canonical SET.  Unlike the previous methods, which made use of the symmetries of the variations and expansions, there is no mandate that the GW SET derived here is symmetric. In fact, this is the case for electrodynamics (see Appendix~\ref{App:Belinfante}).
\fi

Equation~\eqref{eq:GRd_SET} can now be applied to GR to find the canonical SET. In this setting, the fields are the metric perturbation themselves, which propagate on a flat background, and thus
\begin{equation}
\label{eq:g-exp-Noeth}
g_{\alpha _\beta} = \eta_{\alpha \beta} + h_{\alpha \beta}\,,
\end{equation}
with $h_{\alpha \beta} \sim \order{\epsilon}$. The Lagrangian must thus be expanded in the fields to the appropriate order.  Any quantity that depends on the curvature of the background metric vanishes since the latter is the Minkowski metric.  Expanding the Lagrangian to first order leaves odd terms in Eq.~\eqref{eq:GRd_SET}, which vanish upon short-wavelength averaging.  Expanding the Lagrangian in Eq.~(\ref{GR_Action}) to second order about a Minkowski background yields
\begin{align}
\label{eq:GRd_Lag2O}
\mathcal{L} &= \frac{1}{64 \pi G}\left[\partial_\alpha h \, \partial^\alpha h + 2 \partial_\alpha h_{\beta \gamma} \, \partial^\beta h^{\alpha \gamma} \right. \nonumber \\
&\left. - 2\partial^\alpha h \, \partial_\beta {h^\beta}_\alpha - \partial_\gamma h_{\alpha \beta} \, \partial^\gamma h^{\alpha \beta}\right]\,.
\end{align}

With this Lagrangian density, we can now find the equations of motion for the metric perturbation and the GW SET. The former can be computed from the Euler-Lagrange equations:
\begin{equation}
\label{eq:GRd_FE}
\frac{\partial \mathcal{L}}{\partial h_{\alpha \beta}}-\partial_\gamma \frac{\partial \mathcal{L}}{\partial(\partial_\gamma h_{\alpha \beta})}=\frac{1}{2}\partial_\gamma \partial^\gamma h^\TT_{\alpha \beta}=0\,,
\end{equation}
where we have imposed the TT gauge after variation of the Lagrangian. Using Eq.~\eqref{eq:GRd_Lag2O} in Eq.~\eqref{eq:GRd_SET}, short-wavelength averaging, integrating by parts and using the TT gauge condition and Eq.~\eqref{eq:GRd_FE}, one finds
\begin{equation}
{\Theta^\alpha}_\beta = \frac{1}{32 \pi G} \left \langle\partial^\alpha h^{TT}_{\gamma \delta} \, \partial_\beta h^{\gamma \delta}_{TT}\right \rangle.
\end{equation}
This expression is identical to all others found in this section.

\subsection{Derivation of Physical Quantities: $\dot{E}$ and $\dot{P}$}
\label{sec:GR_E}

Ultimately, one is interested in calculating physical, observable quantities from the GW SET that can be measured at spatial infinity, $\iota^0$.  Two examples are the rate of energy and linear momentum transported by GWs away from any system per unit time 
\begin{subequations}
\begin{align}
\label{eq:Edot}
\dot{E} &= -\oint_\infty \Theta^{0 i}d^2S_i\,, \\
\label{eq:Pdot}
\dot{P}^i &= - \oint_\infty \Theta^{ij}d^2S_j\,,
\end{align}
\end{subequations}
where $\Theta^{\alpha i}$ is the $(\alpha,i)$ component of the GW SET.  

These observables can be simplified through the shortwave approximation, which assumes the characteristic wavelength of radiation $\lambda_{c}$ is much shorter than the observer's distance to the center of mass $r$, i.e.~the observer is in the so-called \emph{far-away wave zone} so that $r \gg \lambda_{c}$. When this is true, the propagating fields can be expanded as~\cite{PW}
\be
\phi_{L} = \frac{\lambda_{c}}{r} f_{1,L}(\tau) + \left(\frac{\lambda_{c}}{r}\right)^{2} f_{2,L}(\tau) + {\cal{O}}\left[\left(\frac{\lambda_{c}}{r}\right)^{3}\right]\,,
\ee
where $\tau = t - r/v$ is retarded time and $v$ is the speed of propagation of the field. Moreover, the spacetime (partial) derivative of the field then satisfies
\begin{equation}
\label{eq:ShortwaveSimplification}
\partial_\alpha \phi_{L} = - \frac{1}{v}\,k_\alpha \,\partial_\tau \phi_{L} + \order{\frac{\lambda_{c}^2}{r^2}}\,,
\end{equation}
where $k^\alpha$ is a unit normal 4-vector normal to the $r={\rm{const}}$ surface $\left(k_\alpha \equiv \left(-1,N_i\right)\right)$ with $N_i={x_i}/{r}$ and $x^i$ are spatial coordinates on the 2-sphere.

With this at hand, we can now simplify the observable quantities $\dot{E}$ and $\dot{P}^{i}$ in GR. The rate at which energy and linear momentum are removed from the system is
\begin{subequations}
\label{eq:GR_Edot1}
\begin{align}
\dot{E}_\GR &= -\frac{R^2}{32 \pi G} \oint \avg{\dot{h}^\TT_{\gamma \delta} \; \partial^i h^{\gamma \delta}_\TT}d^2S_i\,,
\\
\dot{P}_\GR^{i} &= -\frac{R^2}{32 \pi G} \oint \avg{\partial^{i}{h}^\TT_{\gamma \delta} \; \partial^j h^{\gamma \delta}_\TT}d^2S_j\,,
\end{align}
\end{subequations}
where $\dot{h}_{\alpha \beta}$ is the partial derivative of $h_{\alpha \beta}$ with respect to coordinate time $t$ and $d^2S_i = R^2N_i\,d\Omega$.  Incorporating the shortwave approximation into Eq.~\eqref{eq:GR_Edot1} gives
\begin{subequations}
\label{eq:GR_Edot2}
\begin{align}
\dot{E}_\GR &= -\frac{R^2}{32 \pi G} \int \avg{\partial_\tau h^\TT_{\gamma \delta} \; \partial_\tau h^{\gamma \delta}_\TT}d\Omega\,,
\\
\label{eq:GRe_LinearMomentum}
\dot{P}^i &= -\frac{R^2}{32 \pi G} \int \avg{N^i \, \partial_\tau h^\TT_{\gamma \delta} \, \partial_\tau h^{\gamma \delta}_\TT} d \Omega\,,
\end{align}
\end{subequations}
which are the final expressions we seek for the GW SET-related observables. To proceed further, one would have to specify a particular physical system, solve the field equations for the metric perturbation for that given system, and then insert the solution in the above equations to carry out the integral; all of this is system specific and outside the scope of this paper. 

\section{Scalar-Tensor Theory}
\label{sec:ST}

Following the procedure presented in the previous section, we now derive the GW SET in scalar-tensor theories, focusing only in the theory proposed by Jordan, Fierz, Brans, and Dicke as an example~\cite{Fierz:1956zz,Jordon,Brans:1961sx}. This theory was developed in an attempt to satisfy Mach's principle and its action is 
\begin{equation}
\label{ST_Action}
S_{\ST}=\frac{1}{16 \pi G} \int d^4x \sqrt{-g}\left(\Phi R - \omega \frac{\nabla_\alpha \Phi \, \nabla^\alpha \Phi}{\Phi}\right)\,,
\end{equation}
where $R$ is the Ricci scalar, $\Phi$ is the scalar field, and $\omega$ is a coupling constant.  In the GR limit, the scalar field $\Phi$ becomes unity.  The field equations for scalar-tensor theory are
\begin{align}
\label{eq:ST_FieldEqn1}
X_{\alpha \beta} &= G_{\alpha \beta} + \frac{1}{\Phi}\left(\nabla_\alpha \nabla_\beta \Phi - g_{\alpha \beta} \, \nabla_\gamma \nabla^\gamma \Phi\right) \nonumber \\
&+ \frac{\omega}{\Phi}\left(\nabla_\alpha \Phi \, \nabla_\beta \Phi - \frac{1}{2} g_{\alpha \beta} \, \nabla_\gamma \Phi \, \nabla^\gamma \Phi\right)=0\,,
\end{align}
and
\begin{equation}
\label{eq:ST_PhiEoM}
Y_{\alpha \beta} = \nabla_\gamma \nabla^\gamma \Phi = 0 \,,
\end{equation}
where we vary the action with respect to the metric and scalar field respectively.  Lastly, we will make use of the ``reduced field"~\cite{Will:1993ns} (see Appendix~\ref{App:ReducedField} for derivation of this quantity),
\begin{equation}
\label{eq:ReducedFieldDef}
\theta_{\alpha \beta} = h_{\alpha \beta} - \frac{1}{2}\tilde{g}_{\alpha \beta}\, h - \frac{1}{\tilde\Phi}\tilde{g}_{\alpha \beta}\,\varphi \,,
\end{equation}
where $\tilde{g}_{\alpha \beta}$ is the background metric and $\tilde\Phi$ is a background scalar field.  As we will show, the field equation for the reduced field is simply
\begin{equation}
\label{eq:ThetaEoM}
\square \theta^\TT_{\alpha \beta} = 0 \,
\end{equation}
in vacuum.

\subsection{Perturbed Action Method}
\label{sec:STa}

Let us begin by decomposing the metric as in Eq.~\eqref{eq:GRa_LinGrav} and the scalar field via 
\begin{equation}
\label{eq:phi-exp}
\Phi = \tilde{\Phi} + \epsilon \; \varphi\,,
\end{equation}
where $\varphi = \order{\epsilon}$.  The action will decompose in a similar manner to that of Eq.~\eqref{eq:GR_Action-exp}, where the expanded action terms are
\begin{widetext}
\begin{subequations}
	\label{eq:STa_ActionExpansion}
	\begin{align}
	S^{\tiny (0)}_{\ST} &=\dfrac{1}{16 \pi G} \int d^4x \; \sqrt{-\tilde{g}} \, \tilde{\Phi}\left(\tilde{R} - \frac{\omega}{\tilde{\Phi}} \tilde{\nabla}_\alpha \tilde{\Phi} \tilde{\nabla}^\alpha \tilde{\Phi}\right)\, ,
	\\
	\label{eq:STa_ActionExpansion-1}
	S^{\tiny (1)}_{\ST} &=-\dfrac{1}{16 \pi G} \int d^4x \sqrt{-\tilde{g}} \, \tilde{\Phi}\left[\tilde{R}^{\alpha \beta}\theta_{\alpha \beta} - \frac{2}{\tilde{\Phi}} \left(6+2\omega \right) \tilde{\square}\varphi- \left(1+\omega\right)\tn_\alpha \tn_\beta \theta^{\alpha \beta} - \frac{1}{2} \tilde{\square} \theta\right] \, ,
	\\
	S^{\tiny (2)}_{\ST}&=-\dfrac{1}{16 \pi G} \int d^4x \sqrt{-\tilde{g}}\, \tilde{\Phi}\left[\frac{1}{\tilde{\Phi}^2}\tilde{R}\,\varphi^2+\frac{1}{\tilde{\Phi}}\tilde{R}^{\alpha \beta}\theta_{\alpha \beta}\,\varphi + \frac{1}{4}\tilde{R}\,\theta^{\alpha \beta}\theta_{\alpha \beta} -\tilde{R}^{\alpha \beta} \, {\theta_\alpha}^\gamma \theta_{\beta \gamma} - \frac{1}{8}\tilde{R}\,{\theta}^2 \right. \nonumber \\ 
	&\left. + \frac{1}{2}\tilde{R}^{\alpha \beta}\,\theta\,\theta_{\alpha \beta} + \frac{1}{\phi_0}\left(\frac{3}{2}+\omega\right)\tilde{\nabla}_\alpha \varphi \, \tilde{\nabla}^\alpha \varphi + \frac{1}{4}\tilde{\nabla}_\gamma \theta_{\alpha \beta} \, \tilde{\nabla}^\gamma \theta^{\alpha \beta} -\frac{1}{8}\tilde{\nabla}_\alpha \theta \, \tilde{\nabla}^\alpha \theta - \frac{1}{2}\tilde{\nabla}_\beta \theta_{\alpha \gamma} \, \tilde{\nabla}^{\gamma} \theta^{\alpha \beta}\right]\,.
	\end{align}
\end{subequations}
\end{widetext}
All trace terms have been taken with respect to the background metric.  As argued before, the $S^{(1)}_\ST$ term will not contribute to the GW SET upon variation, leaving us with the effective action 
\begin{equation}
\label{eq:STa_EffAct}
S_{\ST}^{\tiny \mbox{eff}} = S^{(0) \, \tiny \mbox{eff}}_\ST + \epsilon^2 S^{(2) \, \tiny \mbox{eff}}_\ST + \order{\epsilon^3}\,.
\end{equation}

The independent variation of the effective action with respect to $\theta^{\alpha \beta}$ and $\varphi$  gives the first order field equations. In particular, variation with respect to $\theta^{\alpha \beta}$ gives
\begin{equation}
\label{eq:STa_ThetaEoM1}
\tilde{\square}\theta_{\alpha \beta} - 2\tn^\gamma \tn_{(\alpha} \theta_{\beta) \gamma} - \frac{1}{2}\tilde{g}_{\alpha \beta} \theta = 0\,,
\end{equation}
while variation with respect to $\varphi$ yields
\begin{equation}
\label{eq:STa_VarphiEoM1}
\tilde{\square}\varphi=0 \,.
\end{equation}
Notice that Eq.~\eqref{eq:STa_ThetaEoM1} is identical to the first-order Einstein equations in Eq.~\eqref{eq:GRa_EoM} with the replacement $h_{\alpha \beta} \to \theta_{\alpha \beta}$.  Therefore, by utilizing the TT gauge condition on $\theta_{\alpha \beta}$, namely
\begin{subequations}
\label{eq:TT-gauge-ST}
\begin{align}
\tn^\alpha \theta_{\alpha \beta} ^\TT &= 0 \,, \\
\theta^\TT &= 0 \,,
\end{align}
\end{subequations}
the equation of motion for the reduced field $\theta_{\alpha \beta}^\TT$ is
\begin{equation}
\label{eq:STa_ThetaEoMFin}
\tilde{\square} \theta^\TT_{\alpha \beta} = 0\,.
\end{equation}

Variation of the effective action with respect to the background metric 
\begin{equation}
\delta S^{(2) \tiny \mbox{eff}}_\ST = \int d^4x \sqrt{-\tilde{g}} \, t_{\alpha \beta}^\ST \, \delta \tilde{g}^{\alpha \beta}\,
\end{equation}
gives the GW SET
\begin{equation}
\label{eq:STa_GWSETDef}
\Theta_{\alpha \beta}^\ST = -2 \avg{t_{\alpha \beta}^\ST}\,.
\end{equation}
Carrying out the variation, we find
\begin{align}
\label{eq:STa_PreSET}
\Theta^\ST_{\alpha \beta} &= \frac{1}{32 \pi G} \tilde{\Phi} \avg{\tn_\alpha \theta^{\mu \nu} \, \tn_\beta \theta_{\mu \nu} + \frac{6+4\omega}{\tilde{\Phi}^2}\tn_\alpha \varphi \, \tn_\beta \varphi \right. \nonumber \\
& \left. -\frac{1}{2} \tn_\alpha \theta \, \tn_\beta \theta + \tn_\gamma \theta \, \tn^\gamma \theta_{\alpha \beta} + 4 \tn_{[\gamma}\theta_{\delta] \beta} \, \tn^\delta {\theta_\alpha}^\gamma \right. \nonumber \\
& \left. + \tilde{g}_{\alpha \beta}\left(\tn_\gamma \theta_{\delta \sigma} \, \tn^\sigma \theta^{\gamma \delta} + \frac{1}{4} \, \tn_\gamma \theta \, \tn^\gamma \theta \right. \right. \nonumber \\
& \left. \left. - \frac{1}{2} \, \tn_\gamma \theta_{\delta \sigma} \, \tn^\gamma \theta^{\delta \sigma} - (3+2\omega)\tn_\gamma \varphi \, \tn^\gamma \varphi \right)}\,.
\end{align}
At this point we may impose the TT gauge condition and use integration by parts to simplify Eq.~\eqref{eq:STa_PreSET} with the use of Eqs.~\eqref{eq:STa_VarphiEoM1} and~\eqref{eq:STa_ThetaEoMFin}.  The resulting GW SET is
\begin{equation}
\label{eq:STa_FinalSET}
\Theta^\ST_{\alpha \beta}=\frac{\tilde{\Phi}}{32 \pi G} \left \langle \tilde{\nabla}_{\alpha}\theta^{\gamma \epsilon}_{\TT} \, \tilde{\nabla}_{\beta} \theta^{\TT}_{\gamma \epsilon} + \frac{1}{\tilde{\Phi}^2} \left(6+4\omega\right)\tilde{\nabla}_{\alpha}\varphi \, \tilde{\nabla}_{\beta} \varphi \right \rangle \,.
\end{equation}
This result is consistent with that found in~\cite{Will:1993ns} for $\tilde{g}_{\alpha \beta} \to \eta_{\alpha \beta}$.
\subsection{Perturbed Field Equation Method}
\label{sec:STb}
Let us begin by expanding the metric as in Eq.~\eqref{eq:GRb_gExp} and similarly expand the scalar field as\begin{equation}
\Phi = \phi_0 + \varphi^{(1)} + \varphi^{(2)}\,,
\end{equation}
where $\phi_0$ is now the constant background field. 

The expansion of the field equations to first-order yields the equations of motion for the fields $h_{\alpha \beta}^{(1)}$ and $\varphi^{(1)}$ 
\begin{align}
\label{eq:STb_FE-order1}
\frac{\phi_0}{2} \, \partial_\gamma \partial^\gamma \theta^{\tiny (1)}_{\alpha \beta}  &= \phi_0 \, \partial_\gamma \partial_{(\alpha} \theta^{\tiny (1)}_{\beta)}{}^{\gamma} - \frac{\phi_0}{2} \eta_{\alpha \beta} \, \partial_\gamma \partial_\delta \theta^{\tiny (1) \gamma \delta}\,,
\\
 \partial_\gamma \partial^\gamma \varphi^{(1)} &= 0\,.
\end{align}
Working in the TT gauge as in the previous subsection [see Eq.~\eqref{eq:TT-gauge-ST}] confirms our vacuum field equation [Eq.~\eqref{eq:ThetaEoM}] in a Minkowski background.

As in Eq.~\eqref{eq:GRb_FE-order2}, the GW SET will be defined as the average of the expansion of the field equations to quadratic order in the linear perturbation terms:
\begin{align}
\label{eq:STb_FE-order2}
8 \pi \Theta^\ST_{\alpha \beta} \equiv - \avg{X_{\alpha \beta}\left[\left(h^{\tiny (1)}\right)^{2}, \left(\varphi^{\tiny (1)}\right)^2, h^{\tiny (1)}\varphi^{\tiny (1)}\right]} \,,
\end{align}
where we recall $X_{\alpha \beta}$ is given by Eq.~\eqref{eq:ST_FieldEqn1}.
The details of the expansion of the field equations to second order in the perturbations will be omitted for brevity, but the procedure follows that in Sec.~\ref{sec:GRb}.  Once this is done, we use Eq.~\eqref{eq:STb_FE-order2} to find the GW SET
\begin{align}
\label{eq:STb_SET}
\Theta^\ST_{\alpha \beta}=\frac{\phi_0}{32 \pi G} \avg{\partial_{\alpha}\theta^{\gamma \epsilon}_{\TT} \, \partial_{\beta} \theta^{\TT}_{\gamma \epsilon} + \frac{1}{\phi_0^2} \left(6+4\omega\right)\partial_{\alpha}\varphi \, \partial_{\beta} \varphi}\,.
\end{align}
We arrive at this equation using integration by parts, imposing the TT gauge condition, Eqs.~\eqref{eq:ST_PhiEoM} and~\eqref{eq:ThetaEoM} for a Minkowski background. This GW SET agrees with that in Sec.~\ref{sec:STa} for a flat background.

\subsection{Landau-Lifshitz Method}
\label{sec:STc}

The Landau-Lifshitz method requires the use of the $\mathcal{H}^{\alpha \mu \beta \nu}$ tensor density or a variation of it to obtain a conservation law of the form of Eq.~\eqref{eq:GRc_Hconserv} in Sec.~\ref{sec:GRc}. In principle, one could use the same $\mathcal{H}^{\alpha \mu \beta \nu}$ tensor density as that used in GR [see Eq.~\eqref{eq:GRc_HTensor}], and one would obtain the same rate of energy and linear momentum loss in a binary system~\cite{Lee:1974pt}. But in practice, it is easier to use an improved $\mathcal{H}^{\alpha \mu \beta \nu}$ tensor density that simplifies the calculations of the modified Landau-Lifshitz pseudo-tensor $t^{\alpha \beta}_{\LL,\ST}$. Following the work of \cite{Nutku:1969ApJ}, we choose
\begin{equation}
\label{eq:STc_HTensor}
\mathcal{H}^{\alpha \mu \beta \nu} = \Phi^2 \left(\textgoth{g}^{\alpha \beta}\textgoth{g}^{\mu \nu}- \textgoth{g}^{\alpha \mu}\textgoth{g}^{\beta \nu}\right)\,.
\end{equation}
This satisfies the relation
\begin{equation}
\label{eq:STc_HTensor2}
\partial_\mu \partial_\nu \mathcal{H}^{\alpha \mu \beta \nu} = 2 \left(-g\right) \Phi^2 \left(X^{\alpha \beta} + \frac{8 \pi}{\Phi}t^{\alpha \beta}_{\LL,\ST}\right)\,,
\end{equation}
where $X^{\alpha \beta} = 0$ are the field equations [Eq.~\eqref{eq:ST_FieldEqn1}].
When the GR limit is taken, this agrees exactly with what was found in Sec.~\ref{sec:GRc}.  The final term in Eq.~\eqref{eq:STc_HTensor2} will be called the scalar-tensor pseudo-tensor and is given by~\cite{Nutku:1969ApJ}
\begin{align}
\label{eq:STc_tLLN}
t^{\alpha \beta}_{\LL,\ST} &= \Phi t^{\mu \nu}_{\LL}  + \frac{1}{8 \pi} \left[ 2 \Phi^{,(\alpha} \Gamma^{\beta)}_{\gamma \delta}g^{\gamma \delta} - 2 g^{\gamma (\alpha} \Gamma^{\beta)}_{\gamma \delta} \Phi^{,\delta} \right. \nonumber \\
&\left. + g^{\alpha \beta}(2 \Phi^{,\gamma} \Gamma^\delta_{\gamma \delta} - \Phi_{, \sigma} \Gamma^{\sigma}_{\gamma \delta}g^{\gamma \delta})  - 2 \Phi^{,(\alpha}g^{\beta) \gamma} \Gamma^\delta_{\gamma \delta} \right. \nonumber \\
&\left. + g^{\mu \alpha}g^{\nu \beta} \Phi_{,\gamma} \Gamma^\gamma_{\mu \nu} \right] \nonumber \\
&+ \frac{1}{16 \pi \Phi} \left[2 (\omega -1) \Phi^{,\alpha}\Phi^{,\beta} + (2 - \omega)g^{\alpha \beta} \Phi_{,\gamma}\Phi^{,\gamma}\right]\,.
\end{align}
One could write this expression entirely in terms of the gothic metric, but this does not simplify the resulting expression. 

The GW SET can now be obtained through Eq.~\eqref{eq:GRc_ThetaEqn} after expanding Eq.~\eqref{eq:STc_HTensor2} to leading non-vanishing order. Doing so, making use of the TT gauge, integrating by parts, and using the field equations for the linear perturbations, the final GW SET is found to be
\begin{equation}
\label{eq:STc_SET}
\Theta^\ST_{\alpha \beta}=\frac{\phi_0}{32 \pi G} \avg{\partial_{\alpha}\theta^{\gamma \epsilon}_{\TT} \, \partial_{\beta} \theta^{\TT}_{\gamma \epsilon} + \frac{1}{\phi_0^2} \left(6+4\omega\right)\partial_{\alpha}\varphi \, \partial_{\beta} \varphi}\,.
\end{equation}
Equation~\eqref{eq:STc_SET} is identical to the GW SETs previously found.

\subsection{Noether Current Method}
\label{sec:STd}
The derivation of the canonical SET relies on Eq.~\eqref{eq:GRd_SET}, which requires we expand the Lagrangian to second order through the metric decomposition of Eq.~\eqref{eq:g-exp-Noeth} and the field decomposition of Eq.~\eqref{eq:phi-exp}. Doing so, we find
\begin{align}
\mathcal{L}_{\ST}&=\frac{\phi_0}{32 \pi G} \left(\frac{1}{4}\partial^\alpha \theta \, \partial_\alpha \theta+\partial_\beta \theta_{\alpha \gamma} \, \partial^\gamma \theta^{\alpha \beta} \right. \nonumber \\
&\left. -\frac{1}{2}\partial_\gamma \theta_{\alpha \beta} \, \partial^\gamma \theta^{\alpha \beta}-\frac{\left(3+2\omega \right)}{\phi_0^2}\partial_\alpha \varphi \, \partial^\alpha \varphi\right)\,.
\end{align}

With this second order Lagrangian at hand, we can now follow the same steps as in Sec.~\ref{sec:GRd} to calculate the canonical SET. The first step is to ensure the Euler-Lagrange equations are satisfied.  For the reduced field $\theta_{\alpha \beta}$, the Euler-Lagrange equations give
\begin{equation}
\label{STd_ELtheta}
\partial_\gamma \partial^\gamma \theta^\TT_{\alpha \beta}=0\,,
\end{equation}
where we have used the TT gauge after varying the Lagrangian.  As for $\varphi$, the Euler-Lagrange equations lead to
\begin{equation}
\label{eq:STd_ELphi}
\partial_\gamma \partial^\gamma \varphi =0\,.
\end{equation}
The next step is to compute the Noether's current.  Summing over the variation of all fields, we find
\begin{align}
\label{eq:STd_Current1}
{j^\alpha}_\beta = \left( -\frac{\partial \mathcal{L}}{\partial \left(\partial_\alpha \theta_{\mu \nu}\right)} \partial_\beta \theta_{\mu \nu} -\frac{\partial \mathcal{L}}{\partial \left(\partial_\alpha \varphi\right)} \partial_\beta \varphi + {\eta^\alpha}_\beta \mathcal{L} \right).
\end{align}
The current in Eq.~\eqref{eq:STd_Current1} gives a pseudo-tensor that is not gauge invariant, but after short-wavelength averaging, these terms vanish. Using the TT gauge condition and imposing the first-order field equations, one then finds
\begin{align}
\label{eq:STd_SET}
\Theta^\ST_{\alpha\beta}=\frac{\phi_0}{32 \pi G} \avg{\partial_{\alpha}\theta^{\TT}_{\gamma \delta} \, \partial_{\beta} \theta^{\gamma \delta}_{\TT} + \frac{1}{\phi_0^2} \left(6+4\omega\right)\partial_{\alpha}\varphi \, \partial_{\beta} \varphi}.
\end{align}
The result here is consistent with those presented in the previous sections.

\subsection{Derivation of Physical Quantities: $\dot{E}$ and $\dot{P}$}

The four different ways of deriving the GW SET in scalar-tensor theory all produced the same result, and hence, one can use any of them to derive $\dot{E}$ and $\dot{P}$. Inserting the GW SET into Eqs.~\eqref{eq:Edot} and~\eqref{eq:Pdot}, the rate of change of energy and linear momentum carried away by all propagating degrees of freedom is
\begin{subequations}
\begin{align}
\label{eq:ST_Edot}
\dot{E}_\ST &= -\frac{\phi_0\,R^2}{32 \pi G} \int \avg{\partial_\tau \theta^\TT_{\gamma \delta} \, \partial_\tau \theta^{\gamma \delta}_\TT 
+ \frac{6+4\omega}{\phi_0^2} \partial_\tau \varphi \, \partial_\tau \varphi}d\Omega\,,
\\
\dot{P}^i_\ST &= -\frac{\phi_0\,R^2}{32 \pi G} \int N^i\avg{\partial_\tau \theta^\TT_{\gamma \delta} \, \partial_\tau \theta^{\gamma \delta}_\TT 
 + \frac{6+4\omega}{\phi_0^2} \partial_\tau \varphi \, \partial_\tau \varphi}d\Omega\,.
\end{align}
\end{subequations}
In the GR limit, we find that the rate of energy and momentum loss is identical to that of Sec.~\ref{sec:GR_E}.  

\section{Einstein-\AE{}ther Theory}
\label{sec:EA}
In this section, we study Einstein-\AE{}ther theory~\cite{Jacobson:2000xp} by following the work of Foster~\cite{Foster:2006az}. We begin with the action
\begin{align}
\label{eq:EA_Action}
S_\EA &= \frac{1}{16 \pi G} \int d^4x \sqrt{-g}\left(R  \right. \nonumber \\
&\left.- {K^{\alpha \beta}}_{\gamma \delta} \, \nabla_\alpha u^\gamma \nabla_\beta u^\delta + \lambda \left(u^\alpha u_\alpha +1\right)\right)\,,
\end{align}
where 
\begin{align}
\label{eq:EA_Keqn}
{K^{\alpha \beta}}_{\gamma \epsilon} = c_1g^{\alpha \beta}g_{\gamma \epsilon}+c_2{\delta^\alpha}_\gamma{\delta^\beta}_\epsilon + c_3{\delta^\alpha}_\epsilon{\delta^\beta}_\gamma - c_4u^\alpha u^\beta g_{\gamma \epsilon}\,.
\end{align}
The vector $u^\alpha$ in Eq.~\eqref{eq:EA_Action} is the \AE{}ther field, which is unit timelike due to the Lagrange multiplier $\lambda$ constraint. The quantities $c_{i}$ are coupling constants of the theory. Certain combinations of these constants will typically appear in the perturbed field equations, namely
\begin{subequations}
	\begin{align}
	c_{14} &= c_1+c_4  \,, \\
	c_\pm &= c_1 \pm c_3 \,, \\
	c_{123} &= c_1+c_2+c_3 \,.
	\end{align}
\end{subequations} 

Varying the action with respect to the metric tensor and the \AE{}ther vector field yields the field equations of the theory in vacuum: 
\begin{align}
\label{eq:VT_FE1}
G_{\alpha \beta} &= S_{\alpha \beta}\,,
\\
\label{eq:VT_FE2}
\nabla_\beta {K^\beta}_\alpha &= -\lambda u_\alpha - c_4\left(u_\beta \nabla^\beta u^\gamma \right) \nabla_\alpha u_\gamma\,,
\end{align}
where $G_{\alpha \beta}$ is the Einstein tensor and $S_{\alpha \beta}$ is the \AE{}ther contribution to the field equations given by
\begin{align}
S_{\alpha \beta} &= \nabla_\gamma \left({K^\gamma}_{(\alpha} u_{\beta)} + K_{(\alpha \beta)} u^\gamma - {K_{(\alpha}}^\gamma u_{\beta)}\right) \nonumber \\
& + c_1 \left(\nabla_\alpha u_\gamma \, \nabla_\beta u^\gamma - \nabla_\gamma u_\alpha \, \nabla^\gamma u_\beta\right) \nonumber \\
& + c_4 \left(u^\gamma  u^\delta \, \nabla_\gamma u_\alpha \, \nabla_\delta u_\beta\right) + \lambda \, u_\alpha u_\beta \nonumber \\
& - \frac{1}{2}g_{\alpha \beta}\left({K^\gamma}_\delta \, \nabla_\gamma u^\delta\right)\,,
\end{align}
where
\begin{equation}
{K^\alpha}_\gamma = {K^{\alpha \beta}}_{\gamma \delta} \nabla_\beta u^\delta\,.
\end{equation}
Variation of the action with respect to the Lagrange multiplier $\lambda$ gives the constraint,
\begin{equation}
\label{eq:VT_FElambda}
u^\alpha u_\alpha = -1\,.
\end{equation}
Contracting Eq.~\eqref{eq:VT_FE2} with $u^{\alpha}$, we can solve for $\lambda$ to obtain
\begin{equation}
\lambda = u^\alpha \nabla_\beta {K^\beta}_\alpha + c_4 \left(u^\alpha \nabla_\alpha u^\beta \right) \left(u^\gamma \nabla_\gamma u_\beta \right)\,.
\end{equation}

Unlike in Secs.~\ref{sec:GR} and~\ref{sec:ST}, we will use an irreducible decomposition of all fields rather than a single reduced field. This will have the effect of cleanly separating the independent modes of propagation. Expanding the metric as in Eq.~\eqref{eq:GRa_LinGrav}, the \AE{}ther field can be decomposed via
\begin{equation}
\label{eq:EA_uDecomp}
u^\alpha = t^\alpha + \omega^\alpha\,,
\end{equation}
where $|\omega^\alpha| = \order{h}$ and $t^\alpha=(-1,0,0,0)$ and it is a timelike unit vector with respect to the background metric $\tilde{g}_{\alpha \beta}$, i.e.~$\tilde{g}_{\alpha \beta} t^\alpha t^\beta = -1$.  We next decompose the metric perturbation into  tensor, vector, and scalar modes as follows
\begin{equation}
h^{\alpha \beta} = t^\alpha t^\beta h_{00} + 2 \mathcal{P}^{(\alpha}_i t^{\beta)}h^{0i} + \mathcal{P}^{\alpha}_i \mathcal{P}^\beta_j h^{ij} \,,
\end{equation}
where $\mathcal{P}^\alpha_i$ is the background spatial projector
\begin{equation}
\mathcal{P}^{\alpha \beta} = \tilde{g}^{\alpha \beta} + t^\alpha t^\beta\,.
\end{equation}
The \AE{}ther, vector and tensor perturbations are further decomposed into their transverse and longitudinal parts,
\begin{subequations}
\label{eq:VT_PerturbationParts}
\begin{align}
\omega^\alpha &= t^\alpha \omega_0 + \mathcal{P}^\alpha_i \left(\nu^i + \partial^i \nu \right) \,, \\
h^{0i} &= \gamma^i + \partial^i \gamma \,, \\
h^{ij} &= \phi^{ij}_\TT + \frac{1}{2}P_{ij}\left[f\right] + 2 \partial_{(i} \phi_{j)} + \partial_i \partial_j \phi \,,
\end{align}
\end{subequations}
with $P_{ij}[f] \equiv \tilde{g}_{ij}\,\tilde{\nabla}_k \tilde{\nabla}^k f - \tilde{\nabla}_i \tilde{\nabla}_j f$, $\phi^{ij}_\TT$ a TT spatial tensor and $\nu^i{}_{,i}=\gamma^i{}_{,i}= \phi^i{}_{,i}=0$.  

The degrees of freedom of the theory, the \AE{}ther field $u^{\alpha}$ and the metric tensor $g_{\alpha \beta}$, have thus been replaced by their decompositions $(\omega_{0},\nu^{i},\nu)$ and $(h^{00},\gamma^{i},\gamma,\phi_{ij}^{\TT},f,\phi^{i},\phi)$, but these quantities can be further constrained.  The timelike condition on the \AE{}ther field requires that
\begin{equation}
\label{eq:w0_constraint}
\omega_0 = -\frac{1}{2}h_{00}\,,
\end{equation}
to leading order in the perturbations. We also choose  the gauge conditions
\begin{subequations}
	\label{eq:EA_GaugeConditions}
\begin{align}
\partial_i \omega^i &= 0 \,, \\
\partial_i h^{0i}&= 0 \,, \\
\partial_i \partial_{[j}{h_{k]}}^i &=0 \,,
\end{align}
\end{subequations}
or equivalently $\nu = \gamma = \phi_i =0$.

\subsection{Perturbed Action Method}
\label{sec:EAa}
The metric is decomposed as in Eq.~\eqref{eq:GRa_LinGrav} while the \AE{}ther field takes the form of Eq.~\eqref{eq:EA_uDecomp} with $\omega^\alpha$ replaced by $\epsilon \, \omega^\alpha$.  We expand the action to second-order (see App.~\ref{App:EA_Lag}) and find the effective action 
\begin{equation}
\label{eq:VTa_EffAct}
S_{\VT}^{\tiny \mbox{eff}} = S^{(0) \, \tiny \mbox{eff}}_\VT + \epsilon^2 S^{(2) \, \tiny \mbox{eff}}_\VT + \order{\epsilon^3}\,.
\end{equation}
Recall that the $S^{(1) \tiny \mbox{eff}}_\VT$ is not important for our consideration in this paper due to the angular averaging.  

We start by solving for the first-order equations of motions for the fields.  After expanding the action, we decompose the perturbations into the various transverse and longitudinal parts given in Eq.~\eqref{eq:VT_PerturbationParts}.  Starting with the tensor mode $\phi^{ij}_\TT$,
\begin{equation}
\label{eq:EAa_tensorEoM}
\frac{\delta S^{\tiny (2) \tiny \mbox{eff}}_\EA}{\delta \phi^{ij}_\TT} = (1-c_+)\tilde{\partial}_0^2\phi^\TT_{ij} - \tilde{\triangle}\phi^\TT_{ij}=0\,,
\end{equation}
where we have focused on terms quadratic in the tensor mode from the action.  The $\tilde{\triangle}$ in Eq.~\eqref{eq:EAa_tensorEoM} is the spatial Laplacian operator associated with the background spacetime $\tilde{\triangle} = \tilde{g}^{ij} \tn_i\tn_j$, while the $\tilde{\partial}_0$ operator is the time derivative in this background $\tilde{\partial}_0 = t^{\alpha} \tn_{\alpha}$.  We may rewrite Eq.~\eqref{eq:EAa_tensorEoM} in a more compact form using a modified wave equation,
\begin{equation}
\label{eq:EAa_TensorEoM}
\tilde{\square}_2 \phi^\TT_{ij}=0\,,
\end{equation}
where the wave operator is defined as
\begin{equation}
\tilde{\square}_2 \equiv (1-c_+)\tilde{\partial}_0^2 - \tilde{\triangle} \,.
\end{equation}
From this, we are able to see that the wave speed for the tensor mode is 
\begin{equation}
v_{\tiny \mbox{T}}^2 \equiv \frac{1}{1-c_+}\,.
\end{equation}

We perform the same procedure to the vector modes $\gamma^i$ and $\nu^i$.  The variations become
\begin{subequations}
\label{eq:EAa_VectorEoM1}
\begin{align}
\frac{\delta S^{\tiny (2) \tiny \mbox{eff}}_\EA}{\delta \gamma^i} &= c_{14}\tilde{\partial}^2_0\left(\gamma_i+\nu_i\right)+\frac{1}{2}\tilde{\triangle}\left[\left(1-c_-\right)\gamma_i-c_-\nu_i\right] =0\,,\\
\frac{\delta S^{\tiny (2) \tiny \mbox{eff}}_\EA}{\delta \nu^i} &= c_{14}\tilde{\partial}^2_0\left(\gamma_i+\nu_i\right) - \frac{1}{2}\tilde{\triangle}\left[c_- \gamma_i + 2c_1 \nu_i\right]=0\,.
\end{align}
\end{subequations}
We can combine these equations to give us a relation between the variables $\gamma_i$ and $\nu_i$,
\begin{equation}
\label{eq:EAa_VectorModeRelation}
\gamma_i = -c_+ \nu_i\,.
\end{equation}
Equipped with this relation, we are able to solve for another wave equation, this time for the vector mode,
\begin{align}
\label{eq:EAa_VectorEoM2}
\tilde{\square}_1 \nu_i &\equiv \frac{2\left(1-c_+\right)c_{14}}{2c_1-c_+c_-}\tilde{\partial}^2_0\nu_i - \tilde{\triangle}\nu_i \nonumber \\
&= 0\,.
\end{align}
The wave speed for the vector mode can be read off as
\begin{equation}
v_{\tiny \mbox{V}}^2 \equiv \frac{2c_1-c_+c_-}{2(1-c_+)c_{14}}\,.
\end{equation}

Lastly, we investigate the scalar modes $h_{00}$, $\phi$, and $F$. The relevant variations are
\begin{subequations}
\label{eq:ScalarEoMs}
\begin{align}
\frac{\delta S^{\tiny (2) \tiny \mbox{eff}}_\EA}{\delta h_{00}} &= \frac{1}{32\pi G}\tilde{\triangle} \left(c_{14}\,h_{00}-F\right)\,,\\
\frac{\delta S^{\tiny (2) \tiny \mbox{eff}}_\EA}{\delta \phi} &= \frac{1}{32\pi G}\,\partial_0\,  \partial_0\,\tilde{\triangle} \left[\left(1+c_2\right)F+c_{123}\tilde{\triangle} \phi\right]\,,\\
\label{eq:partc}
\frac{\delta S^{\tiny (2) \tiny \mbox{eff}}_\EA}{\delta F} &= \frac{1}{32\pi G}\left[\tilde{\triangle}h_{00}+\frac{1}{2}\left(1+c_++2c_2\right)\tilde{\partial}^2_0F\right. \nonumber \\
&\left.+\left(1+c_2\right)\tilde{\triangle}\tilde{\partial}^2_0\phi-\frac{1}{2}\tilde{\triangle}F\right]\,,
\end{align}
\end{subequations}
with $F\equiv \tn_i \tn^i f$.
From the vanishing of the first two scalar mode variations, we are able to find the relations
\begin{subequations}
	\label{eq:EAa_ScalarRelations}
	\begin{align}
		h_{00} &=\frac{1}{c_{14}}F \,,\\
		\tilde{\triangle}\phi &= -\frac{1+c_2}{c_{123}}F \,.
	\end{align}
\end{subequations}
Using Eq.~\eqref{eq:EAa_ScalarRelations} in the final variation, Eq.~\eqref{eq:partc}, we are able to find the modified wave equation for the scalar mode
\begin{align}
\label{eq:EAa_ScalarEoM}
\tilde{\square}_0F &\equiv \frac{(1-c_+)(2+2c_2+c_{123})c_{14}}{(2-c_{14})c_{123}}\tilde{\partial}^2_0F-\tilde{\triangle}F \nonumber \\
&= 0\,,
\end{align}
with a propagation speed of
\begin{equation}
v_{\tiny \mbox{S}}^2 \equiv \frac{(2-c_{14})c_{123}}{(1-c_+)(2+2c_2+c_{123})c_{14}}\,.
\end{equation}

We know from the previous sections that we can find the GW SET by using \begin{equation}
\delta S^{(2) \tiny \mbox{eff}}_\VT = \int d^4x \sqrt{-\tilde{g}} \, t_{\alpha \beta}^\VT \, \delta \tilde{g}^{\alpha \beta}\,,
\end{equation}
where we define the GW SET to be
\begin{equation}
\label{eq:VTa_GWSETDef}
\Theta_{\alpha \beta}^\VT = -2 \avg{t_{\alpha \beta}^\VT}\,.
\end{equation}
To simplify the result, we decompose the resulting GW SET into its tensor $(T)$, vector $(V)$, and scalar $(S)$ pieces.  In addition to this, we make use of the equations of motion given in Eqs.~\eqref{eq:EAa_TensorEoM},~\eqref{eq:EAa_VectorEoM2} and~\eqref{eq:EAa_ScalarEoM}.  The GW SET is
\begin{widetext}
\begin{subequations}
\begin{align}
(T):  \;\Theta^\VT_{\alpha \beta} &=\frac{1}{32 \pi G} \avg{\tilde{\nabla}_\alpha \phi^{ij}_{\TT} \tilde{\nabla}_\beta \phi^\TT_{ij}} \,, \\
(V):  \;\Theta^\VT_{\alpha \beta} &= \frac{1-c_+}{16 \pi G}\avg{\left(2c_1-c_+c_-\right)\tn_\alpha \nu^i \, \tn_\beta \nu_i + c_+\left(\frac{c_+(2c_1-c_+c_-)-2c_{14}}{2c_1-c_+c_-}\right)\,t_\alpha t_\beta \left(\tilde{\partial}^2_0\nu^i\right) \nu_i}\,, \\
(S): \;\Theta^\VT_{\alpha \beta} &= \frac{1}{64 \pi G} \avg{\left(\frac{2-c_{14}}{c_{14}}\right)\tn_\alpha F \,\tn_\beta F \right. \nonumber \\
	&\left. - \frac{2}{c_{14}}t_\alpha t_\beta\left(\frac{2c_{14}-3c_2+2c_{14}c_2-c_++2c_2c_+}{c_{123}}\left(\tilde{\partial}^2_0 F\right)+\frac{3c_1-2(c_++c_{14})}{c_{14}}\tilde{\triangle} F\right)F}\,.
\end{align}
\end{subequations}
\end{widetext}
Note that in this situation, the GW SET has terms explicitly dependent on the $t^\alpha$, the Lorentz-violating background \AE{}ther field. However, we will see that these terms do not affect physical observables.

\subsection{Perturbed Field Equation Method}
\label{sec:EAb}
We begin by expanding the metric as in Eq.~\eqref{eq:GRb_gExp} and expanding the \AE{}ther field as
\begin{equation}
u^\alpha = t^\alpha + \omega^{(1) \alpha} + \omega^{(2) \alpha}\,.
\end{equation}
We next expand the field equations in Eqs.~\eqref{eq:VT_FE1} and~\eqref{eq:VT_FE2} to $\order{h}$ in order to find the equations of motion for the perturbed fields
\begin{widetext}
\begin{subequations}
\begin{align}
\label{eq:VTb_FEa}
G_{\alpha \beta}^{(1)} - S_{\alpha \beta}^{(1)} &= \frac{1}{2}\left[2\partial_\gamma \partial_{(\alpha}{h_{\beta)}^{(1)\gamma}} - \partial_\alpha \partial_\beta h^{(1)} - \partial_\gamma \partial^\gamma h^{(1)}_{\alpha \beta} + g_{\alpha \beta} \left(\partial_\gamma \partial^\gamma h^{(1)} - \partial^\gamma \partial^\delta h^{(1)}_{\gamma \delta}\right)\right] \nonumber \\
& +c_1 \left[t^\gamma t_{(\alpha} \, \partial_{\beta)} \partial_\delta h_\gamma^{(1) \delta} - \partial_\gamma \partial^\gamma \left(t^\delta t_{(\alpha} h_{\beta) \delta}^{(1)}\right) - \frac{1}{2} \ddot{h}_{\alpha \beta}^{(1)} + t_{(\alpha} \partial_{\beta)} \partial_\gamma \omega^{(1) \gamma} - \partial_{(\alpha}\dot{\omega}_{\beta)}^{(1)} \right. \nonumber \\
& \left. - \partial_\gamma \partial^\gamma \left(t_{(\alpha}\omega_{\beta)}^{(1)}\right)\right] - c_2 \, g_{\alpha \beta} \left[\partial_\gamma \dot{\omega}^{(1) \gamma} - \frac{1}{2} \ddot{h}^{(1)}\right] -c_3 \left[t^\gamma t_{(\alpha} \, \partial_{\beta)} \partial_\delta h_\gamma^{(1) \delta} - \partial_\gamma \partial^\gamma \left(t^\delta t_{(\alpha} h_{\beta) \delta}^{(1)}\right) \right. \nonumber \\
&\left. + \frac{1}{2} \ddot{h}_{\alpha \beta}^{(1)} + t_{(\alpha} \partial_{\beta)} \partial_\gamma \omega^{(1) \gamma} + \partial_{(\alpha}\dot{\omega}_{\beta)}^{(1)}- \partial_\gamma \partial^\gamma \left(t_{(\alpha}\omega_{\beta)}^{(1)}\right)\right] \nonumber \\
& +c_4 \left[\frac{1}{2}\, t_\alpha t_\beta \left(t^\gamma t^\delta \, \partial_\epsilon \partial^\epsilon h^{(1)}_{\gamma \delta} - 2 t^\gamma \partial_\delta \dot{h}_{\gamma}^{(1) \delta}\right)+2 \, t^\gamma t_{(\alpha}\ddot{h}_{\beta) \gamma}^{(1)} - t^\gamma t^\delta t_{(\alpha} \partial_{\beta)}\dot{h}_{\gamma \delta}^{(1)}+ 2\, t_{(\alpha} \ddot{\omega}_{\beta)}^{(1)} - t_\alpha t_\beta  \partial_\gamma \dot{\omega}^{(1) \gamma}\right]\,,\\
\label{eq:VTb_FEb}
\left(\nabla_\beta {K^\beta}_\alpha\right)^{(1)} &= c_1 \left[\frac{1}{2} t^\beta \partial_\gamma \partial^\gamma h_{\alpha \beta}^{(1)} - t^\beta \partial_\gamma \partial_{[\alpha}h_{\beta]}^{(1) \gamma} + \partial_\beta \partial^\beta \omega_\alpha^{(1)}\right] + c_2 \left[\frac{1}{2} \, \partial_\alpha \dot{h}^{(1)} + \partial_\alpha \partial_\beta \omega^{(1) \beta}\right] \nonumber \\
& - c_3 \left[\frac{1}{2} t^\beta \partial_\gamma \partial^\gamma h_{\alpha \beta}^{(1)} - t^\beta \partial_\gamma \partial_{(\alpha}h_{\beta)}^{(1) \gamma} - \partial_\alpha \partial_\beta \omega^{(1)\beta}\right] + c_4 \left[\frac{1}{2} t^\beta t^\gamma \partial_\alpha \dot{h}_{\beta \gamma}^{(1)} - t^\beta \ddot{h}_{\alpha \beta}^{(1)} - \ddot{\omega}_\alpha^{(1)}\right]\,.
\end{align}
\end{subequations}
\end{widetext}
We are now able to decompose the first-order field equations into the modes of propagation.  For the tensor mode,
\begin{equation}
\label{eq:EAb_TensorFE}
\partial_\gamma \partial^\gamma \phi_{i j}^\TT + c_+ \ddot{\phi}^\TT_{ij}=0\,.
\end{equation}
We notice that Eq.~\eqref{eq:EAb_TensorFE} may be written to leading order as a modified wave equation,
\begin{equation}
\label{eq:TensorEoM}
\square_2 \phi^\TT_{ij} = 0\,,
\end{equation}
which matches what we found in Eq.~\eqref{eq:EAa_TensorEoM} with the background taken to be Minkowski.  The vector modes are found by contracting Eqs.~\eqref{eq:VTb_FEa} and ~\eqref{eq:VTb_FEb} with the projector $\mathcal{P}^\alpha_i$.  The results are,
\begin{subequations}
\label{eq:VTb_VectorModeFE}
\begin{align}
\label{eq:VTb_VectorModeFE1}
c_{14}\left(\ddot{\gamma}_i+\ddot{\nu}_i\right)+\frac{1}{2}\, \partial_j \partial^j \left[\left(1-c_-\right)\gamma_i-c_-\nu_i\right] &=0\,,\\
\label{eq:VTb_VectorModeFE2}
c_{14}\left(\ddot{\gamma}_i+\ddot{\nu}_i\right) - \frac{1}{2}\, \partial_j \partial^j\left[c_- \gamma_i + 2c_1 \nu_i\right]&=0\,.
\end{align}
\end{subequations}
Equations~\eqref{eq:VTb_VectorModeFE1} and~\eqref{eq:VTb_VectorModeFE2} are then solved simultaneously to generate the relation $\gamma_i = -c_+ \nu_i$, leading to the modified wave equation,
\begin{equation}
\square_1 \nu_i \equiv \frac{2\left(1-c_+\right)c_{14}}{2c_1-c_+c_-}\ddot{\nu}_i - \partial_j \partial^j \nu_i=0\,,
\end{equation}
which matches Eq.~\eqref{eq:EAa_VectorEoM2} for a Minkowski background.  The scalar mode equation of motion is found by looking at the scalar field equations: $G^{(1)}_{ii}-S^{(1)}_{ii}=0$, $G^{(1)}_{00}-S^{(1)}_{00}=0$, and $\left(\nabla_\beta {K^\beta}_i \right)^{(1)}=0$, namely
\begin{subequations}
\label{eq:VTb_ScalarModeFE}
\begin{align}
\label{eq:VTb_ScalarModeFE1}
&-\partial_j \partial^j h_{00}- \frac{1}{2}\left(2+2c_2+c_{123}\right)\ddot{F} \nonumber \\
& +\frac{1}{2} \partial_j \partial^j F - \frac{1}{2}\left(2+2c_2+c_{123}\right) \partial_j \partial^j \ddot{\phi}= 0 \,, \\
\label{eq:VTb_ScalarModeFE2}
& \frac{1}{2} \left(c_{14} \, h_{00} + F \right) =0 \,, \\
\label{eq:VTb_ScalarModeFE3}
& \frac{1}{2} \partial_0 \partial_i \left(c_{123} \partial_j \partial^j \phi + c_{14} \, h_{00} + c_2 \, F\right) = 0\,.
\end{align}
\end{subequations}
From Eqs.~\eqref{eq:VTb_ScalarModeFE2} and~\eqref{eq:VTb_ScalarModeFE3} we find the relations 
\begin{subequations}
\begin{align}
h_{00} &= -\frac{1}{c_{14}}F \,, \\
\partial_i \partial^i \, \phi &= - \frac{1+c_2}{c_{123}}F \,.
\end{align}
\end{subequations}
These combined with Eq.~\eqref{eq:VTb_ScalarModeFE1} give the modified wave equations for the scalar mode
\begin{equation}
\square_0 F \equiv \frac{(1-c_+)(2+2c_2+c_{123})c_{14}}{(2-c_{14})c_{123}}\ddot{F}-\partial_j \partial^j F=0\,,
\end{equation}
which agrees with Eq.~\eqref{eq:EAa_ScalarEoM} for a Minkowski background.

We now expand the field equations to $\order{h^2}$ to find the GW SET.  Retaining only the tensor mode ($T$) terms gives
\begin{align}
\Theta_{\alpha \beta}^{(T)} &= \frac{1}{32 \pi G} \avg{\partial_\alpha \phi^\TT_{ij} \partial_\beta \phi^{ij}_\TT - 2\,\square_2 \,\phi_{\alpha i}^\TT \, \phi_\beta^{\TT \, i} \right. \nonumber \\
	& \left. - \frac{1}{2}\eta_{\alpha \beta} \, \square_2 \, \phi^{\TT}_{ij} \, \phi_\TT^{ij}+ t_\alpha t_\beta \,c_+ \ddot{\phi}_{ij}^\TT \phi_{\TT}^{ij} } \,.
\end{align}
We use Eq.~\eqref{eq:TensorEoM} to simplify this expression to
\begin{equation}
\label{eq:TensorSET1}
\Theta_{\alpha \beta}^{(T)} = \frac{1}{32 \pi G}\avg{\partial_\alpha \phi^\TT_{ij} \partial_\beta \phi^{ij}_\TT + t_\alpha t_\beta \,c_+ \ddot{\phi}_{ij}^\TT \phi_{\TT}^{ij}}\,.
\end{equation}
This procedure is again repeated for the vector ($V$) and scalar ($S$) modes to give additional terms for the GW SET,
\begin{align}
\label{eq:VectorSET1}
\Theta_{\alpha \beta}^{(V)} &= \frac{1-c_+}{16 \pi G}\avg{\left(2c_1-c_+c_-\right)\partial_\alpha \nu^i \partial_\beta \nu_i \right. \nonumber \\
&\left. + t_\alpha t_\beta \left(c_+^2-2c_4(1-c_+)\right)\ddot{\nu}^i \nu_i}\,,
\end{align}
\begin{align}
\label{eq:ScalarSET1}
\Theta_{\alpha \beta}^{(S)} &= \frac{1}{64\pi  G   c_{14}}\avg{\left(2-c_{14}\right)\partial_\alpha F \partial_\beta F \right. \nonumber \\
&\left. -\left(\frac{c_{14}\left(1-c_+\right)\left(2+2c_2+c_{123}\right)}{c_{123}}\right) t_\alpha t_\beta \ddot{F}F}\,.
\end{align}
Note that all of the terms proportional to $t^\alpha$ here differ from those found in Sec~\ref{sec:EAa}.  This is okay, since there is no way to define a unique GW SET. However, as we will see in Sec.~\ref{sec:EAe}, the resulting physical observables will be the same.

\subsection{Landau-Lifshitz Method}
\label{sec:EAc}
We now construct a tensor density $\mathcal{H}^{\alpha \mu \beta \nu}$ to obtain a conservation law of the form of Eq.~\eqref{eq:GRc_Hconserv} in Sec.~\ref{sec:GRc}.  By keeping $\mathcal{H}^{\alpha \mu \beta \nu}$ the same as that found in Eq.~\eqref{eq:GRc_HTensor} (as also done in~\cite{Lee:1974nq} to derive the GW SET for other vector-tensor theories), we obtain Eq.~\eqref{eq:GRc_HProp1} where $G^{\alpha \beta}$ is again the Einstein tensor and $t^{\alpha \beta}_\LL$ is the Landau-Lifshitz pseudotensor found in Eq.~\eqref{eq:GRc_tLL}.  We now substitute the field equations found in Eq.~\eqref{eq:VT_FE1} which gives us (including $T^{\alpha \beta}_\mat$ for completeness)
\begin{align}
\partial_\mu \partial_\nu \mathcal{H}^{\alpha \mu \beta \nu} &= 2 \left(-g\right)\left(8 \pi G T^{\alpha \beta}_\mat+S^{\alpha \beta}\right)+ 16 \pi G \left(-g\right)t^{\alpha \beta}_{\LL} \nonumber \\
&= 16 \pi G \left(-g\right) \left(T^{\alpha \beta}_\mat + \frac{1}{8 \pi G}S^{\alpha \beta} + t^{\alpha \beta}_\LL\right)\nonumber \\
&=16 \pi G \left(-g\right) \left(T^{\alpha \beta}_\mat + \bar{t}^{\alpha \beta}_\LL\right)\,,
\end{align}
where $\bar{t}^{\alpha \beta}_\LL$ is the modified Landau-Lifshitz pseudotensor
\begin{equation}
\label{eq:VT_tLLMod}
\bar{t}^{\alpha \beta}_\LL = \frac{1}{8 \pi G}S^{\alpha \beta} + t^{\alpha \beta}_\LL \,.
\end{equation}
Following the same procedure as in Sec.~\ref{sec:GRc}, we find the GW SET to be
\begin{equation}
\label{eq:VTc_SET}
\Theta^{\alpha \beta}_\VT = \frac{1}{16 \pi G }\avg{\partial_\mu \partial_\nu \mathcal{H}^{\alpha \mu \beta \nu}}\,.
\end{equation}

As before in the Landau-Lifshitz Method sections, we expand the pseudotensor in Eq.~\eqref{eq:VT_tLLMod} to second order in the metric and \AE{}ther perturbations given in Eqs.~\eqref{eq:GRc_gothg2} and~\eqref{eq:EA_uDecomp} respectively.  Once this is done, we impose the decomposition and solve Eq.~\eqref{eq:VTc_SET} for each mode.  Recall that this method provides no way to solve for the field equations for the first order perturbations. We already calculated these equations in the previous section, and so we can use them here to simplify the GW SET. Doing so, tensor part of the GW SET is
\begin{equation}
\label{eq:VTc_TensorSET1}
\Theta^{(T) \, \alpha \beta}_\VT= \frac{1}{32 \pi G}\avg{\partial^\alpha \phi^\TT_{ij} \partial^\beta \phi^{ij}_\TT + t^\alpha t^\beta \,c_+ \ddot{\phi}_{ij}^\TT \phi_{\TT}^{ij}}\,.
\end{equation}
We apply this analysis to the vector and scalar modes and find
\begin{align}
\Theta^{(V) \, \alpha \beta}_\VT &= \frac{1-c_+}{16 \pi G}\avg{\left(2c_1-c_+c_-\right)\partial_\alpha \nu^i \partial_\beta \nu_i \right. \nonumber \\
&\left. + t_\alpha t_\beta \left(c_+^2-2c_4(1-c_+)\right)\ddot{\nu}^i \nu_i}\,,
\end{align}

\begin{align}
\Theta^{(S) \, \alpha \beta}_\VT &= \frac{1}{64\pi  G   c_{14}}\avg{\left(2-c_{14}\right)\partial_\alpha F \partial_\beta F \right. \nonumber \\
&\left. -\left(\frac{c_{14}\left(1-c_+\right)\left(2+2c_2+c_{123}\right)}{c_{123}}\right) t_\alpha t_\beta \ddot{F}F}\,.
\end{align}
Notice that the results in this section match those given in Sec.~\ref{sec:EAb}.  This is because in both cases, the expansion of the field equations was used to find the GW SET.

\subsection{Noether Current Method}
\label{sec:EAd}
The derivation of the canonical SET relies on Eq.~\eqref{eq:GRd_SET}, which requires we expand the Lagrangian density to second order (see App.~\ref{App:EA_Lag}) through the metric decomposition of Eq.~\eqref{eq:g-exp-Noeth} and the \AE{}ther decomposition of Eq.~\eqref{eq:EA_uDecomp}\footnote{See~\cite{Eling:2005zq} for related work on deriving the non-symmetric GW SET in Einstein-\AE ther theory using the Noether current method without applying any perturbations.}. We decompose the results into each tensor, vector, and scalar component before applying the Euler-Lagrange equations to each individual field contribution.  For the tensor mode, we only have one field $\phi^\TT_{ij}$.  The resulting Euler-Lagrange equation is 
\begin{equation}
\label{eq:EAd_tensorEL}
\frac{1}{32\pi}\left[\left(1-c_+\right)\ddot{\phi}^\TT_{ij} - \triangle \phi^\TT_{ij}\right]=0\,.
\end{equation}
This is again identical to the modified wave equation of Eq.~\eqref{eq:EAa_tensorEoM} around a Minkowski background.  The Euler-Lagrange equations for the vector modes are
\begin{subequations}
	\begin{align}
		\frac{c_{14}}{8\pi}\left(\ddot{\gamma}_i+\ddot{\nu}_i\right)+\frac{1}{16\pi}\triangle \left[\left(1-c_-\right)\gamma_i-c_-\nu_i\right] &= 0 \,, \\
		\frac{c_{14}}{8\pi}\left(\ddot{\gamma}_i+\ddot{\nu}_i\right) - \frac{1}{16\pi} \triangle \left(c_- \gamma_i + 2c_1 \nu_i\right) &= 0\,,
	\end{align}
\end{subequations}
for $\gamma_i$ and $\nu_i$ respectively.  These reduce to the same modified field equations for the vector mode around a Minkowski background. Using the relation $\gamma_i = -c_+ \nu_i$, these equations can be combined as
\begin{equation}
\label{eq:EAd_vectorEL}
\square_1 \nu_i=0\,.
\end{equation}
The scalar Euler-Lagrange equations are
\begin{subequations}
\begin{align}
\frac{1}{32\pi} \triangle \left(c_{14}h_{00}-F\right)&=0\,,\\
\frac{1}{32\pi}\partial_\alpha \partial^\alpha \left[c_{123}\, \triangle \phi + \left(1+c_2\right)F\right]&=0\,,\\
\frac{1}{64\pi}\triangle\left[\triangle F - 2 \triangle h_{00} - \left(1+c_++2c_2\right) \ddot{F} \right. &\nonumber \\
\left.- 2\left(1+c_2\right)\triangle \ddot{\phi}\right]&=0\,,
\end{align}
\end{subequations}
when we vary with respect to $h_{00}$, $\phi$, and $f$ respectively.  Combining these equations gives us
\begin{equation}
\label{eq:EAd_scalarEL}
\square_0 F =0\,,
\end{equation}
with the relations from Eq.~\eqref{eq:EAa_ScalarRelations} found explicitly.

With the equations of motion at hand, we may use Eq.~\eqref{eq:GRd_SET} to solve for the GW SET.  We again look at each mode individually and use Eqs.~\eqref{eq:EA_GaugeConditions},~\eqref{eq:EAd_tensorEL},~\eqref{eq:EAd_vectorEL}, and~\eqref{eq:EAd_scalarEL} to simplify the results 
\begin{subequations}
\label{eq:EAd_TensorSET1}
\begin{align}
\Theta_{\alpha \beta}^{(T)} &= \frac{1}{32\pi G} \avg{\partial_\alpha \phi^\TT_{ij} \partial_\beta \phi^{ij}_\TT + c_+ \, t_\alpha \,\partial_\beta \phi^\TT_{jk} \dot{\phi}^{jk}_\TT} \,, \\
\Theta_{\alpha \beta}^{(V)} &= \frac{1-c_+}{32\pi G} \avg{(2c_1-c_+c_-)\partial_\alpha \nu^i \partial_\beta \nu_i \right. \nonumber \\
& \left. + \left(c_+^2-2c_4(1-c_+)\right)t_\alpha \partial_\beta \nu_i\,\dot{\nu}^i}\,,\\
\Theta_{\alpha \beta}^{(S)} &= \frac{1}{64\pi G \,c_{123}c_{14}} \avg{\left(2c_++c_{14}(4+3c_2-c_+) \right. \right. \nonumber \\
& \left. \left. -2c_2 -4\right)\partial_\alpha F \partial_\beta F + \left(2(c_+ - 2)+c_2(3c_{14}c_+-2) \right. \right. \nonumber \\
& \left. \left. +c_{14}(2+c_+^2)\right) t_\alpha \partial_\beta F \dot{F} \right. \nonumber \\
& \left. + 4(1+c_2)(1-c_{14}) \, \mathcal{P}_\alpha^i  \, \partial_\beta F \, \partial_i F}\,.
\end{align}
\end{subequations}
The above SET is indeed conserved, i.e.~$\nabla_{\alpha} (\Theta^{\alpha \beta}_{(T)} + \Theta^{\alpha \beta}_{(V)} + \Theta^{\alpha \beta}_{(S)}) = 0$, and thus, we can calculate the rate of change of the energy and linear momentum carried away to spatial infinity by all propagating modes. If we do so, we indeed find the correct answer, i.e.~the same answer as what one finds with all other methods to compute these quantities. Indeed, Foster~\cite{Foster:2006az} used the Noether charge method, which is related to the Noether method shown here, to find the correct $\dot{E}$. However, the above GW SET is somewhat unsatisfactory because it is clearly not symmetric in its two indices or gauge invariant, and thus, if we tried to calculate $\dot{E}$ by taking the covariant divergence with respect to the second index in the GW SET, we would find the wrong answer.     

To fix these problems, we must apply the Belinfante procedure~\cite{BELINFANTE1940449,Soper:1976bb,Guarrera:2007tu} (App. \ref{App:Belinfante}).  Applying this method to Einstein-\AE ther theory gives the GW SET 
\begin{subequations}
\label{eq:EAd_TensorSET2}
\begin{align}
\Theta_{\alpha \beta}^{(T)} &= \frac{1}{32\pi G} \avg{\partial_\alpha \phi^\TT_{ij} \partial_\beta \phi^{ij}_\TT  + c_+ t_{(\alpha} \partial_{\beta)} \phi^\TT_{jk}\,\dot{\phi}^{jk}_\TT} \,, \\
\Theta_{\alpha \beta}^{(V)} &= \frac{1-c_+}{16\pi G}\avg{\left(2c_1-c_+c_-\right)\partial_\alpha \nu^i \partial_\beta \nu_i \right. \nonumber \\
& \left. + \left(c_+^2-2c_{4}\left(1-c_+\right)\right)t_{(\alpha}\partial_{\beta)} \nu^j\,\dot{\nu_j}}\,, \\
\Theta_{\alpha \beta}^{(S)} &= \frac{1}{64\pi G \,c_{123}c_{14}} \avg{\left(2c_++c_{14}(4+3c_2-c_+) \right. \right. \nonumber \\
& \left. \left. -2c_2 -4\right)\partial_\alpha F \partial_\beta F + \left(2(c_+ - 2)+c_2(3c_{14}c_+-2) \right. \right. \nonumber \\
& \left. \left. +c_{14}(2+c_+^2)\right) t_{(\alpha} \partial_{\beta)} F \dot{F} \right. \nonumber \\
& \left. + 4(1+c_2)(1-c_{14}) \, \mathcal{P}_{(\alpha}^i \, \partial_{\beta)} F \, \partial_i F}\,.
\end{align}
\end{subequations}
Observe that this new GW SET is indeed symmetric, but it is not gauge invariant because of all the terms that are proportional to $t_{(\alpha} \partial_{\beta)}$.  As before, the symmetrized GW SET of  Eq.~\eqref{eq:EAd_TensorSET2} does not match that of Sec.~\ref{sec:EAa} and~\ref{sec:EAb}, but this time the differences actually do propagate into observable quantities. 

Given an actual observation, however, there will be a unique measured value of, for example, the energy flux carried by GWs, so what went wrong? The answer is in the application of the Belinfante procedure~\cite{BELINFANTE1940449,Guarrera:2007tu}. This algorithm is derived assuming the Lagrangian density is invariant under Lorentz transformations. Einstein-\AE{}ther is indeed diffeomorphism invariant, but the solutions of this theory do spontaneously break Lorentz-symmetry. Therefore, when the Einstein-\AE{}ther Lagrangian density is expanded about a Lorentz-violating background solution, it loses its diffeomorphism invariance, and in particular, it loses its Lorentz invariance, making the Belinfante procedure inapplicable. Indeed, we find that all of the differences in observables calculated with the above GW SET are proportional to the Lorentz-violating background \AE{}ther field $t^\alpha$. One could in principle generalize the Belinfante procedure to allow for the construction of SETs in Lorentz-violating theories, but this is beyond the scope of this paper.

\subsection{Derivation of Physical Quantities: $\dot{E}$, $\dot{P}$, and $\dot{L}$}
\label{sec:EAe}
Now that we have acquired the GW SET, we can solve for the physical observables.  The first one we look at is $\dot{E}$, which we use Eq.~\eqref{eq:Edot} to solve for.  The first three methods from Sec.~\ref{sec:EAa},~\ref{sec:EAb} and~\ref{sec:EAc} give $\dot{E}$ as
\begin{align}
\label{eq:EA_Edot1}
\dot{E}_\EA &= -\frac{R^2}{16 \pi G} \oint \avg{\frac{1}{2\,v_{\tiny \mbox{T}}} \, \partial_\tau\phi^{\TT}_{ij} \, \partial_\tau\phi_{\TT}^{ij} \right. \nonumber \\
& \left. + \frac{(1-c_+)(2c_1-c_+c_-)}{v_{\tiny \mbox{V}}} \, \partial_\tau \nu_i \, \partial_\tau \nu^i \right. \nonumber \\
&\left. + \frac{2-c_{14}}{4\,v_{\tiny \mbox{S}}\,c_{14}} \, \partial_\tau F \, \partial_\tau F}\,.
\end{align}
The results here agree with those previously found by Foster~\cite{Foster:2006az} using the Noether charge method~\cite{Wald:1993nt,Iyer:1994ys}, which is different from the Noether current method adopted in this paper.  
Similarly, we can solve for the loss rate of linear momentum.  Using the GW SET from Secs.~\ref{sec:EAa},~\ref{sec:EAb} and~\ref{sec:EAc} we find,
\begin{align}
	\label{eq:EA_Pdot}
	\dot{P}^i_\EA &= -\frac{R^2}{16 \pi G} \int N^i \avg{\frac{1}{2\,v_{\tiny \mbox{T}}} \, \partial_\tau\phi^{\TT}_{ij} \, \partial_\tau\phi_{\TT}^{ij} \right. \nonumber \\
		& \left. + \frac{(1-c_+)(2c_1-c_+c_-)}{v_{\tiny \mbox{V}}} \, \partial_\tau \nu_i \, \partial_\tau \nu^i \right. \nonumber \\
		&\left. + \frac{2-c_{14}}{4\,v_{\tiny \mbox{S}}\,c_{14}} \, \partial_\tau F \, \partial_\tau F}\,.
\end{align}

\section{Conclusion}
\label{conclusion}

In this paper we studied a wide array of methods to calculate the GW SET in theories with propagating scalar, vector and tensor fields.  The methods include the variation of the action with respect to a generic background, the second-order perturbation of the field equations, the calculation of a pseudotensor from the symmetries of a tensor density, and the use of Noether's theorem to derive a canonical GW SET.  Generally, all methods yield the same results, but care should be taken when dealing with theories that break Lorentz symmetry. This is because the procedure that symmetrizes the canonical SET and makes it gauge-invariant (the so-called Belinfante procedure) fails in its standard form in theories that are not Lorentz invariant. In addition to all of this, we present here and for the first time the symmetric GW SET for Einstein-\AE{}ther theory, from which we calculated the rate of energy and linear momentum carried away by all propagating degrees of freedom; the rate of energy loss had been calculated before~\cite{Foster:2006az} and it agrees with the results we obtained. 

The work we presented here opens the door to several possible future studies. One crucial ingredient in binary pulsar and GW observations that has not been calculated here is the rate of angular momentum carried away by all propagating degrees of freedom. The best way to calculate this quantity is through the Landau-Lifshitz pseudo-tensor, but this requires expansions to one higher order in perturbation theory and the careful application of short-wavelength averaging. Once this is calculated, for example in Einstein-\AE{}ther theory, one could combine the result with the rate of energy loss computed in this paper to calculate the rate at which the orbital period and the eccentricity decay in compact binaries. These results could then be used to constrain Einstein-\AE{}ther theory with binary pulsar observations and GW observations. The latter would require the construction of models for the GWs emitted in eccentric inspirals of compact binaries, which in turn requires the energy and angular momentum loss rate. 

Another possible avenue for future work is to calculate the GW SET in more complicated theories, such as TeVeS~\cite{Bekenstein:2004ne} and MOG~\cite{Moffat:2004nw,Moffat:2007ju}. Both of these theories modify Einstein's through the inclusion of a scalar and vector field with non-trivial interactions with the tensor sector and the matter sector. The methods studied in this paper are well-suited for the calculation of the GW SET in these more complicated theories. Once that tensor has been calculated, one could then compute the rate of energy carried away by all propagating degrees of freedom in this theory, and from that, one could compute the rate of orbital period decay in binary systems. Binary pulsar observations and GW observations could then be used to stringently constrain these theories in a new independent way from previous constraints. 

\section*{Acknowledgments}
\label{ackno}

N.Y and A.S.~acknowledge support through the NSF CAREER grant PHY-1250636 and NASA grants NNX16AB98G and 80NSSC17M0041. K.Y. acknowledges support from Simons Foundation and NSF grant PHY-1305682.

\appendix
\section{Brill-Hartle Averaging Scheme}
\label{App:Averaging}
This section will show some of the advantages of the averaging scheme used by Isaacson~\cite{Brill:1964zz}, which he called Brill-Hartle averaging, and we refer to as wavelength-averaging.  The average of some tensor $X^{\alpha \beta}$ will be defined in the following way,
\begin{align}
\label{eq:AppA_AvgDef}
\avg{X^{\alpha \beta}(x)} \equiv & \int  d^4x \, {g^\alpha}_{\alpha^\prime}(x,x^\prime)
 \; {g^\beta}_{\beta^\prime}(x,x^\prime) \nonumber \\
 & \times X^{\alpha^\prime \beta^\prime}(x^\prime) \; f(x,x^\prime)\,,
\end{align}
where ${g^\alpha}_{\alpha^\prime}(x,x^\prime)$ is the bivector of geodesic parallel displacement~\cite{DeWitt:1960fc} that depends on the background geometry and $f(x,x^\prime)$ is the kernel of the integral satisfying
\begin{equation}
\int d^4x f(x,x^\prime) = 1\,.
\end{equation}

There are four useful concepts to consider.  The first will be the commutation of covariant derivatives.  For the tensor $h^{\alpha \beta}$, the commutation of the covariant derivatives is by definition,
\begin{align}
\label{eq:AppA-Commutation}
\avg{\nabla_\gamma \nabla_\delta h^{\alpha \beta}} = \avg{\nabla_\delta \nabla_\gamma h^{\alpha \beta}- {R^\alpha}_{\mu \delta \gamma}h^{\mu \beta} - {R^\beta}_{\mu \delta \gamma}h^{\mu \alpha}} \,,
\end{align}
where ${R^{\alpha}}_{\beta \gamma \delta}$ is the Riemann curvature tensor that corresponds to the background spacetime.  From the field equations, we know that the curvature tensor is sourced by terms of $\order{h^2}$.  Therefore, we can see that the commutation of covariant derivatives goes as
\begin{equation}
\label{eq:AppA_TotalDiv}
\avg{\nabla_\gamma \nabla_\delta h^{\alpha \beta}} = \avg{\nabla_\delta \nabla_\gamma h^{\alpha \beta}} +\order{h^3} \,.
\end{equation}
For our purposes, we neglect these higher order terms, allowing us to commute covariant derivatives without the addition of curvature terms.

The second useful point of emphasis is the vanishing of total divergences,
\begin{equation}
\avg{\nabla_{\mu}X^{\mu\, a_1 \cdots a_{n-1}}}=0\,,
\end{equation}
for some $n$-rank tensor $X^{\mu\, a_1 \cdots a_{n-1}}$ that varies on the scale of the gravitational radiation wavelength.  Substituting this quantity into the averaging integral generates four terms after integration by parts.  The total divergence term will vanish since we turn it into a surface integral.  The remaining three terms will be of $\order{h}$ due to the bivector and kernel varying on scales larger than the wavelengths.  We show this for one term below.  Let $\lambda$ be the scale over which any perturbation varies while $\mathcal{R}$ is the scale length of fluctuations of the background.  We will assume the high-frequency limit, which tells us that $\lambda \ll \mathcal{R}$.  With this in mind, we see how the averaging modifies terms by looking at the scaling:
\begin{align}
\avg{\nabla_{\mu}X^{A}} &\sim {g^\mu}_{\mu^\prime} \cdots {g^{a_{n-1}}}_{{a_{n-1}^\prime}}X^{\mu^\prime \cdots a_{n-1}^\prime} \; \left(\nabla_\mu f \right)\,, \nonumber \\
\avg{\order{\frac{\partial X^A}{\partial \lambda}}} &\sim \order{X^A \;\frac{\partial f}{\partial \mathcal{R}}}\,, \nonumber \\
&\sim \order{\frac{\partial X^A}{\partial \lambda}} \order{\frac{\partial \lambda}{\partial \mathcal{R}}}\,,
\end{align}
where we have simplified notation such that $A = \left(\mu, \alpha_1, \cdots, \alpha_{n-1}\right)$.
We have made use of the fact that $\partial f$ is of $\order{1}$ since it varies on scales related to the background geometry.  Notice the averaging process added a term of $\order{{\lambda}/{\mathcal{R}}}$ to the original tensor being averaged.

The third point to consider is the usefulness of integration by parts, which says that
\begin{equation}
\avg{\nabla_\mu \left(X^{A}\right)Y^B} = -\avg{X^A \left(\nabla_\mu Y^B \right)}\,.
\end{equation}
Notice that the total divergence becomes of $\order{h}$ higher than the quantities remaining.  Due to this, boundary terms do not need to be considered here.

The fourth and final point to consider is that the product of an odd number of the quantities being averaged over will vanish.  This is straightforward when considering oscillatory functions which oscillate with a single frequency.  An integral over an odd number of these quantities vanishes, which is again what happens here due to the fact that the averaging process involves integration.

\section{Electromagnetic Canonical SET}
\label{App:Belinfante}
Consider the Lagrangian for classical electrodynamics in a source free spacetime,
\begin{equation}
	\label{eq:EMlag}
	\mathcal{L} = -\frac{1}{4}F_{\alpha \beta}F^{\alpha \beta}\,,
\end{equation}
where $F_{\alpha \beta} \equiv 2\partial_{[\alpha} A_{\beta]}$, with $A^\alpha$ being the 4-vector potential.  Applying the canonical SET from Eq.~\eqref{eq:GRd_SET} gives,
\begin{equation}
	\label{eq:AppB_CanSET}
	{j^\alpha}_\beta = -\frac{1}{4}{\delta^\alpha}_{\beta} F_{\mu \nu}F^{\mu \nu} - \partial_\beta A_\gamma \; F^{\gamma \alpha}\,.
\end{equation}
Notice that the last term in Eq.~\eqref{eq:AppB_CanSET} is not gauge invariant under the transformation $A^\alpha \rightarrow A^\alpha + \partial^\alpha \epsilon$.  The solution to this is the Belinfante procedure~\cite{BELINFANTE1940449,Soper:1976bb}.

We here use the Guarrera and Hariton~\cite{Guarrera:2007tu} implementation of the Belinfante procedure.  The symmetric tensor is defined as
\begin{align}
	\label{eq:GH_Theta}
	\Theta_\GH^{\alpha \beta} = -\pi_\gamma^{(\alpha} \partial^{\beta)}A^\gamma + \partial_\delta \left(\pi_\gamma^{(\alpha} M^{\beta) \delta}A^\gamma \right) + g^{\alpha \beta}\mathcal{L}\,,
\end{align}
where $\pi_\gamma^{\alpha} \equiv \partial \mathcal L / \partial (\partial_\alpha A^\gamma)$ and $M^{\alpha \beta}$ is the spin tensor for the 4-vector potential.  Applying Eq.~\eqref{eq:GH_Theta} and using the Lorenz gauge and equations of motion in a source free medium,
\begin{subequations}
	\begin{align}
		\partial_\alpha A^\alpha &=0\,, \\
		\partial_\alpha \partial^\alpha A^\beta &=0 \,, 
	\end{align}
\end{subequations}
we arrive at the classical result for the electromagnetic SET
\begin{equation}
	\Theta_\GH^{\alpha \beta} = F^{\alpha \gamma}{F^\beta}_\gamma - \frac{1}{4} g^{\alpha \beta} F^{\gamma \delta}F_{\gamma \delta}\,.
\end{equation}
Equation~\eqref{eq:GH_Theta} was found using the assumption that the theory in question is Lorentz invariant.  This was necessary in order to derive the correct $M^{\alpha \beta}$ tensors, which are the generators of infinitesimal Lorentz transformations.
\section{Derivation of Scalar-Tensor Reduced Field}
\label{App:ReducedField}
In accordance with the procedures outlined in \cite{Will:1993ns}, we consider the notion of a ``reduced field."  The reduced field is the field which enables the field equations to be written as
\be
\label{eq:ST_RedFieldEqn}
\left(-\frac{1}{v_g}\frac{\partial^2}{\partial t^2} + \partial_i \partial^i\right) \mathcal A = - 16 \pi \, \mathcal S\,,
\ee
where $\mathcal A$ is the linear perturbation of the reduced field, $v_g$ is the speed of propagation of the reduced field, and the source $\mathcal S$ is the combination of matter and higher order perturbation effects.  Notice that in GR, the ``reduced field" is the trace-reversed metric perturbation.  

For the work in Sec.~\ref{sec:ST}, the reduced field will need to be derived.  This is accomplished through the use of decomposition.
Let us assume that this decomposition takes the form
\begin{equation}
\label{eq:ST_thetaEqn}
\theta_{\alpha \beta} = h_{\alpha \beta} +C_1\, \tilde{g}_{\alpha \beta}\, h +C_2\, \tilde{g}_{\alpha \beta}\, \varphi,
\end{equation}
where $\tilde{g}_{\alpha \beta}$ in Eq.~\eqref{eq:ST_thetaEqn} is any background metric.  Since we need linear order terms, we expand the action in Eq.~\eqref{ST_Action} to $\order{h}$ and vary with respect to the background metric  to obtain the field equations in terms of $h_{\alpha \beta}$ and $\varphi$, 
\begin{align}
\label{eq:ReducedField_FE-order1}
0 &=-\frac{1}{2}\square \left(h_{\alpha \beta} - \tilde{g}_{\alpha \beta}h -2 \tilde{g}_{\alpha \beta}\frac{\varphi}{\phi_0}\right) \nonumber \\
&+ \frac{1}{2}\left(2\partial_\gamma \partial_{(\alpha}{h_{\beta)}}^\gamma - \partial_\alpha \partial_\beta h - \tilde{g}_{\alpha \beta}\,\partial_\gamma \partial_\delta h^{\gamma \delta} - \frac{2}{\phi_0}\partial_\alpha \partial_\beta \varphi \right)\,.
\end{align}
Any derivative here is with respect to the background metric.  At this point, we substitute in the reduced field using Eq.~\eqref{eq:ST_thetaEqn}
\begin{align}
\label{eq:ReducedField_FE-order1-Expanded}
0 &= -\frac{1}{2}\phi_0\square \left[\theta_{\alpha \beta} + \left(2 \, C_2 - \frac{2}{\phi_0} - \frac{8 C_1 C_2}{1+4C_1}\right)\tilde{g}_{\alpha \beta} \varphi \right. \nonumber \\
& \left. + \left(\frac{2C_1}{1+4C_1} - 1\right)\tilde{g}_{\alpha \beta} \theta\right] + \phi_0 \, \partial_\alpha \partial_\beta \left[\left(\frac{C_1}{1+4C_1}-\frac{1}{2}\right)\theta \right. \nonumber \\
&\left.+\left(C_2-\frac{1}{\phi_0} - \frac{4 C_1 C_2}{1+4C_1}\right)\varphi\right] + \partial_\gamma \partial_{(\alpha}{\theta_{\beta)}}^\gamma \nonumber \\
&+ \frac{\phi_0}{2}\tilde{g}_{\alpha \beta} \partial_\gamma \partial_\delta \theta^{\gamma \delta}\,.
\end{align}
We know we are looking for the only $\square$ term to be $\square \theta_{\alpha \beta}$.  This allows for us to solve for $C_1$ and $C_2$ by eliminating the terms proportional to $\theta$ and $\varphi$ in the first bracket, leading to
\begin{subequations}
	\begin{align}
	C_1 &= -\frac{1}{2},\\
	C_2 &= -\frac{1}{\phi_0}.
	\end{align}
\end{subequations}
With the inclusion of these constants, the linearized reduced field equations become
\begin{equation}
\label{eq:redFE2}
\square \theta_{\alpha \beta} - \frac{2}{\phi_0} \, \partial_\gamma \partial_{(\alpha} {\theta_{\beta)}}^{\gamma} + \tilde{g}_{\alpha \beta} \, \partial_\gamma \partial_\delta \, \theta^{\gamma \delta}=0\,.
\end{equation}
We are now free to impose the Lorenz gauge condition for the reduced field.  This eliminates the final two terms in Eq.~\eqref{eq:redFE2}, leading to the field equations for the reduced field.

\section{Expanded Action for Einstein-\AE{}ther}
\label{App:EA_Lag}
Here, we state the expanded action for Einstein-\AE{}ther. These results were obtained through the use of the xTensor package for Mathematica~\cite{MartinGarcia:2008qz,Brizuela:2008ra}.
\begin{widetext}
\begin{subequations}
	\label{eq:EAa_ActionExpansion}
	\begin{align}
S^{\tiny (0)}_\EA &= \frac{1}{16 \pi G} \int d^4x \sqrt{-\tilde{g}} \left(\tilde{R} - c_1(\tn_\alpha \tu_\beta) (\tn^\alpha \tu^\beta) - c_2 (\tn_\alpha \tu^\alpha) (\tn_\beta \tn^\beta) - c_3 (\tn_\alpha \tu_\beta) (\tn^\beta \tu^\alpha) + c_4 \,\tu^\alpha\, \tilde{u}^\beta\, (\tn_\alpha \tu^\gamma) (\tn_\beta \tu_\gamma)\right)\,, \\
S^{\tiny (1)}_\EA &= \frac{1}{16\pi G} \int d^4x \sqrt{-\tilde{g}}\left[\frac{1}{2}\, h \, \tilde{R} - h^{\alpha \beta} \tilde{R}_{\alpha \beta} + \tilde{\nabla}_\alpha \tilde{\nabla}_\beta h^{\alpha \beta}- \tilde{\square}\,h+ c_1\left(h_{\alpha \gamma} (\tn^\alpha \tu^\beta)(\tn^\gamma \tu_\beta)+2\,\tu^\alpha(\tn_{[\beta}h_{\alpha] \gamma})(\tn^\gamma \tu^\beta) \right. \right. \nonumber \\
& \left. \left. -\frac{1}{2}\,h(\tn_\beta \tu_\alpha)(\tn^\beta \tu^\alpha)- h_{\alpha \gamma}(\tn_\beta \tu^\gamma)(\tn^\beta \tu^\alpha)- \tu^\alpha (\tn_\gamma h_{\alpha \beta})(\tn^\gamma \tu^\beta) - 2(\tn_\alpha \omega_\beta)(\tn^\alpha \tu^\beta)\right) \right. \nonumber \\
&\left. -c_2 \left(\tu^\alpha (\tn_\alpha h)(\tn_\beta \tu^\beta)+ \frac{1}{2}\,h (\tn_\alpha \tu^\alpha)(\tn_\beta \tu^\beta)- 2\,(\tn_\alpha \omega^\alpha)(\tn_\beta \tu^\beta)\right) \right. \nonumber \\
& \left. +c_3\left(2\, \tu^\alpha (\tn_{[\gamma}h_{\beta]\alpha})(\tn^\gamma \tu^\beta)-\tu^\alpha (\tn_\alpha h_{\beta \gamma})(\tn^\gamma \tu^\beta) -\frac{1}{2}\,h(\tn_\alpha \tu_\beta)(\tn^\beta \tu^\alpha) - 2\,(\tn_\alpha \omega_\beta)(\tn^\beta \tu^\alpha)\right) \right. \nonumber \\
&\left. + c_4\left(\frac{1}{2}\,h \,\tu^\alpha \, \tu^\beta (\tn_\alpha \tu^\gamma)(\tn_\beta \tu_\gamma)+h_{\gamma \delta} \, \tu^\alpha \, \tu^\beta(\tn_\alpha \tu^\gamma)(\tn_\beta \tu^\delta)+2\, \tu^\alpha \, \tu^\beta \, \tu^\gamma (\tn_\alpha \tu^\delta)(\tn_\gamma h_{\beta \delta}) \right. \right. \nonumber \\
&\left. \left.- \tu^\alpha \, \tu^\beta \, \tu^\gamma (\tn_\alpha \tu^\delta)(\tn_\delta h_{\beta \gamma}) +2\, \omega^\alpha \, \tu^\beta(\tn_\alpha \tu^\gamma)(\tn_\beta \tu_\gamma)+2 \, \tu^\alpha \, \tu^\beta (\tn_\alpha \omega_\gamma)(\tn_\beta \tu^\gamma)\right)\right]\,, \\
S^{\tiny (2)}_\EA &= \frac{1}{16 \pi G} \int d^4x \sqrt{-\tilde{g}}\left[{h_\alpha}^\beta \, h^{\alpha \gamma} \, \tilde{R}_{\beta \gamma} - \frac{1}{2} \, h \, h^{\alpha \beta} \tilde{R}_{\alpha \beta}+\frac{1}{8} \, h^2 \, \tilde{R}+ \frac{1}{4}(\tn_\alpha h) (\tn^\alpha h) - \frac{1}{2}(\tn_\alpha h)(\tn_\beta h^{\alpha \beta})\right. \nonumber \\
&\left.+\frac{1}{2}(\tn_\alpha h_{\beta \gamma})(\tn^\gamma h^{\alpha \beta}) - \frac{1}{4}(\tn_\alpha h_{\beta \gamma})(\tn^\alpha h^{\beta \gamma}) + c_1\left(\frac{1}{4} \, h_{\alpha \gamma}\,h_{\beta \delta}(\tn^\beta \tu^\alpha)(\tn^\delta \tu^\gamma) - \frac{1}{4} \, \tu^\alpha \, \tu^\beta (\tn_\alpha h^{\gamma \delta})(\tn_\beta h_{\gamma \delta})\right. \right. \nonumber \\
&\left. \left.+\frac{1}{4} \, h_{\alpha \beta}\,h^{\alpha \beta}(\tn_\gamma \tu_\delta)(\tn^\gamma \tu^\delta) - \frac{1}{8} \, h^2(\tn_\alpha \tu_\beta)(\tn^\alpha \tu^\beta) \frac{1}{2}\,h\,h_{\alpha \beta}(\tn_\gamma \tu^\beta)(\tn^\gamma \tu^\alpha) - h_{\alpha \beta}\,{h_\gamma}^\beta(\tn^\alpha \tu^\delta)(\tn^\beta \tu_\delta)\right. \right. \nonumber \\
&\left. \left. +\frac{1}{2}\,h\,h_{\alpha \beta}(\tn^\alpha \tu^\gamma)(\tn^\beta \tu_\gamma) - h\, \tu^\alpha(\tn_{(\alpha} h_{\gamma) \beta})(\tn^\gamma \tu^\beta)+2\,{h_\alpha}^\beta\,\tu^\gamma(\tn_{[\gamma} h_{\beta] \delta})(\tn^\alpha \tu^\delta)+\frac{1}{2}h\, \tu^\alpha(\tn_\beta h_{\alpha \gamma})(\tn^\gamma \tu^\beta) \right. \right. \nonumber \\
&\left. \left. + {h_\alpha}^\beta \, \tu^\gamma(\tn^\alpha \tu^\delta)(\tn_\beta h_{\gamma \delta}) + \tu^\alpha \, \tu^\beta (\tn_{[\gamma}h_{\delta] \beta})(\tn^\delta {h_\alpha}^\gamma) - h (\tn_\alpha \omega_\beta)(\tn^\alpha \tu^\beta) - 2\, h_{\alpha \beta}(\tn_\gamma \omega^\beta)(\tn^\gamma \tu^\alpha) \right. \right. \nonumber \\
&\left. \left. + 2\,\omega^\alpha(\tn_{[\beta}h_{\alpha] \gamma})(\tn^\gamma \tu^\beta) - \omega^\alpha (\tn_\beta h_{\alpha \gamma})(\tn^\beta \tu^\gamma)+2\,h_{\alpha \beta}(\tn^\alpha \tu^\gamma)(\tn^\beta \omega_\gamma)+2\,\tu^\alpha(\tn_{[\beta}h_{\alpha] \gamma})(\tn^\gamma \omega^\beta) \right. \right. \nonumber \\
&\left. \left. - \tu^\alpha(\tn_\beta h_{\alpha \gamma})(\tn^\beta \omega^\gamma) - (\tn_\alpha \omega_\beta)(\tn^\alpha \omega^\beta)\right)+c_2\left(\frac{1}{4}\,h_{\alpha \beta}\,h^{\alpha \beta}(\tn_\gamma \tu^\gamma)(\tn_\delta \tu^\delta)-\frac{1}{4}\,\tu^\alpha\,\tu^\beta(\tn_\alpha h)(\tn_\beta h)\right. \right. \nonumber \\
&\left. \left. +h^{\alpha \beta}\,\tu^\gamma(\tn_\gamma h_{\alpha \beta})(\tn_\delta \tu^\delta)-\frac{1}{2}\,h\,\tu^\alpha (\tn_\alpha h)(\tn_\beta \tu^\beta) - \frac{1}{8}\,h^2(\tn_\alpha \tu^\alpha)(\tn_\beta \tu^\beta)-\tu^\alpha (\tn_\alpha h)(\tn_\beta \omega^\beta)-h(\tn_\alpha \tu^\alpha)(\tn_\beta \omega^\beta) \right. \right. \nonumber \\
&\left. \left. -\omega^\alpha(\tn_\alpha h)(\tn_\beta \tu^\beta)-(\tn_\alpha \omega^\alpha)(\tn_\beta \omega^\beta)\right) +c_3\left(\frac{1}{4}\,h_{\alpha \beta}\, h^{\alpha \beta}(\tn_\gamma \tu_\delta)(\tn^\delta \tu^\gamma)-\frac{1}{4}\,\tu^\alpha \, \tu^\beta (\tn_\alpha h^{\gamma \delta})(\tn_\beta h_{\gamma \delta}) \right. \right. \nonumber \\
&\left. \left. -\frac{1}{8}\,h^2(\tn_\alpha \tu_\beta)(\tn^\beta \tu^\alpha)-h\,\tu^\alpha (\tn_{(\alpha}h_{\beta) \gamma})(\tn^\gamma \tu^\beta)-\frac{1}{2}\,h\,u^\alpha(\tn_\gamma h_{\alpha \beta})(\tn^\gamma \tu^\beta) +2\,{h_\alpha}^\beta\,\tu^\gamma(\tn_{(\gamma}h_{\delta)\beta})(\tn^\alpha \tu^\delta) \right. \right. \nonumber \\
&\left. \left. -{h_\alpha}^\beta\, \tu^\gamma(\tn_\beta h_{\gamma \delta})(\tn^\alpha \tu^\delta) + \tu^\alpha \, \tu^\beta (\tn_{[\gamma}h_{\delta] \beta})(\tn^\gamma {h_\alpha}^\delta) + 2\, \omega^\alpha (\tn_{[\beta}h_{\gamma] \alpha})(\tn^\beta \tu^\gamma) - \omega^\alpha (\tn_\alpha h_{\beta \gamma})(\tn^\beta \tu^\gamma) \right. \right. \nonumber \\
&\left. \left. + 2\tu^\alpha(\tn_{[\beta}h_{\alpha] \gamma})(\tn^\beta \omega^\gamma) - \tu^\alpha(\tn_\beta h_{\alpha \gamma})(\tn^\gamma \omega^\beta)-h(\tn_\alpha \omega_\beta)(\tn^\beta \tu^\alpha) - (\tn_\alpha \omega_\beta)(\tn^\beta \omega^\alpha)\right) \right. \nonumber \\
&\left. + c_4 \left( \frac{1}{4}\, \tu^\alpha\,\tu^\beta\,\tu^\gamma\,\tu^\delta(\tn_\epsilon h_{\alpha \beta})(\tn^\epsilon h_{\gamma \delta}) +2\,\tu^\alpha \, \tu^\beta \, \tu^\gamma \, \tu^\delta (\tn_\beta {h_\alpha}^\epsilon)(\tn_{[\delta}h_{\epsilon] \gamma})- \frac{1}{4}\,h_{\alpha \beta}\,h^{\alpha \beta} \, \tu^\gamma \, \tu^\delta(\tn_\gamma \tu^\epsilon)(\tn_\delta \tu_\epsilon)  \right. \right. \nonumber \\
&\left. \left. + \frac{1}{8}\, h^2 \, \tu^\alpha \, \tu^\beta(\tn_\alpha \tu^\gamma)(\tn_\beta \tu_\gamma)+ \frac{1}{2}\,h\,h_{\alpha \beta} \, \tu^\gamma \, \tu^\beta (\tn_\gamma \tu^\alpha)(\tn_\delta \tu^\beta) + h\, \tu^\alpha \, \tu^\beta \, \tu^\gamma (\tn_\alpha \tu^\delta)(\tn_\gamma h_{\beta \delta}) \right. \right. \nonumber \\
&\left. \left. - \frac{1}{2}\,h\, \tu^\alpha \, \tu^\beta \, \tu^\gamma(\tn_\alpha \tu^\delta)(\tn_\delta h_{\beta \gamma})+2\,h_{\alpha \beta}\, \tu^\gamma \, \omega^\delta (\tn_\gamma \tu^\alpha)(\tn_\delta \tu^\beta) + h \, \tu^\alpha \omega^\beta (\tn_\alpha \tu^\gamma)(\tn_\beta \tu_\gamma) + 4\, \tu^\alpha \, \omega^\beta  (\tn_{(\alpha}\tu^\gamma)(\tn_{\beta)}\omega_\gamma)\right. \right. \nonumber \\
&\left. \left. +2\, h_{\alpha \beta} \, \tu^\gamma \, \tu^\delta (\tn_\gamma \tu^\alpha)(\tn_\delta \omega^\beta) + h\, \tu^\alpha \, \tu^\beta (\tn_\alpha \tu^\gamma)(\tn_\beta \omega_\gamma) + 4\, \tu^\alpha \, \tu^\beta \, \omega^\gamma (\tn_{(\alpha}\tu^\delta)(\tn_{\gamma)}h_{\beta \delta}) + \tu^\alpha \, \tu^\alpha \, \tu^\gamma (\tn_\alpha \omega^\delta)(\tn_\gamma h_{\beta \delta})\right. \right. \nonumber \\
&\left. \left. - \tu^\alpha \, \tu^\beta \omega^\gamma (\tn_\gamma \tu^\delta)(\tn_\delta h_{\alpha \beta})- 4\, \tu^\alpha \, \tu^\beta \, \omega^\gamma (\tn_\alpha \tu^\delta)(\tn_{[\beta}h_{\delta] \gamma}) - \tu^\alpha \, \tu^\beta \, \tu^\gamma (\tn_\alpha \omega^\delta)(\tn_\delta h_{\beta \gamma}) + \tu^\alpha \, \tu^\beta (\tn_\alpha \omega^\gamma)(\tn_\beta \omega_\gamma) \right. \right. \nonumber \\
&\left. \left. + \omega^\alpha \, \omega^\beta (\tn_\alpha \tu^\gamma)(\tn_\beta \tu_\gamma)\right)\right]\,.
	\end{align}
\end{subequations}
\end{widetext}
\bibliography{SET_bib}{}

\begin{thebibliography}{10}

\bibitem{Abbott:2016blz}
B.~P. Abbott {\em et~al.}, ``{Observation of Gravitational Waves from a Binary
  Black Hole Merger},'' {\em Phys. Rev. Lett.}, vol.~116, no.~6, p.~061102,
  2016.

\bibitem{Abbott:2016nmj}
B.~P. Abbott {\em et~al.}, ``{GW151226: Observation of Gravitational Waves from
  a 22-Solar-Mass Binary Black Hole Coalescence},'' {\em Phys. Rev. Lett.},
  vol.~116, no.~24, p.~241103, 2016.

\bibitem{TheLIGOScientific:2016pea}
B.~P. Abbott {\em et~al.}, ``{Binary Black Hole Mergers in the first Advanced
  LIGO Observing Run},'' {\em Phys. Rev.}, vol.~X6, no.~4, p.~041015, 2016.

\bibitem{Abbott:2017vtc}
B.~P. Abbott {\em et~al.}, ``{GW170104: Observation of a 50-Solar-Mass Binary
  Black Hole Coalescence at Redshift 0.2},'' {\em Phys. Rev. Lett.}, vol.~118,
  no.~22, p.~221101, 2017.

\bibitem{Abbott:2017oio}
B.~P. Abbott {\em et~al.}, ``{GW170814: A Three-Detector Observation of
  Gravitational Waves from a Binary Black Hole Coalescence},'' {\em Phys. Rev.
  Lett.}, vol.~119, no.~14, p.~141101, 2017.

\bibitem{TheLIGOScientific:2017qsa}
B.~P. Abbott {\em et~al.}, ``{GW170817: Observation of Gravitational Waves from
  a Binary Neutron Star Inspiral},'' {\em Phys. Rev. Lett.}, vol.~119,
  p.~161101, 2017.

\bibitem{Isaacson:1967zz}
R.~A. Isaacson, ``{Gravitational Radiation in the Limit of High Frequency. I.
  The Linear Approximation and Geometrical Optics},'' {\em Phys. Rev.},
  vol.~166, pp.~1263--1271, 1967.

\bibitem{Isaacson:1968zza}
R.~A. Isaacson, ``{Gravitational Radiation in the Limit of High Frequency. II.
  Nonlinear Terms and the Ef fective Stress Tensor},'' {\em Phys. Rev.},
  vol.~166, pp.~1272--1279, 1968.

\bibitem{Stein:2010pn}
L.~C. Stein and N.~Yunes, ``{Effective Gravitational Wave Stress-energy Tensor
  in Alternative Theories of Gravity},'' {\em Phys. Rev.}, vol.~D83, p.~064038,
  2011.

\bibitem{Landau:1982dva}
L.~D. Landau and E.~M. Lifschits, {\em {The Classical Theory of Fields}},
  vol.~Volume 2 of {\em Course of Theoretical Physics}.
\newblock Oxford: Pergamon Press, 1975.

\bibitem{Noether:1918zz}
E.~Noether, ``{Invariant Variation Problems},'' {\em Gott. Nachr.}, vol.~1918,
  pp.~235--257, 1918.
\newblock [Transp. Theory Statist. Phys.1,186(1971)].

\bibitem{BELINFANTE1940449}
F.~Belinfante, ``On the current and the density of the electric charge, the
  energy, the linear momentum and the angular momentum of arbitrary fields,''
  {\em Physica}, vol.~7, no.~5, pp.~449 -- 474, 1940.

\bibitem{Soper:1976bb}
D.~E. Soper, {\em {Classical Field Theory}}.
\newblock 1976.

\bibitem{Guarrera:2007tu}
D.~Guarrera and A.~J. Hariton, ``{Papapetrou Energy-Momentum Tensor for
  Chern-Simons Modified Gravity},'' {\em Phys. Rev.}, vol.~D76, p.~044011,
  2007.

\bibitem{Jordon}
P.~Jordan, {\em Schwerkraft und Weltall}.
\newblock Friedrich Vieweg and Sohn, Braunschweig, 1955.

\bibitem{Fierz:1956zz}
M.~Fierz, ``{On the physical interpretation of P.Jordan's extended theory of
  gravitation},'' {\em Helv. Phys. Acta}, vol.~29, pp.~128--134, 1956.

\bibitem{Brans:1961sx}
C.~Brans and R.~H. Dicke, ``{Mach's principle and a relativistic theory of
  gravitation},'' {\em Phys. Rev.}, vol.~124, pp.~925--935, 1961.

\bibitem{Jacobson:2000xp}
T.~Jacobson and D.~Mattingly, ``{Gravity with a dynamical preferred frame},''
  {\em Phys. Rev.}, vol.~D64, p.~024028, 2001.

\bibitem{Jacobson:2008aj}
T.~Jacobson, ``{Einstein-aether gravity: A Status report},'' {\em PoS},
  vol.~QG-PH, p.~020, 2007.

\bibitem{Nutku:1969ApJ}
Y.~{Nutku}, ``{The Energy-Momentum Complex in the Brans-Dicke Theory},'' {\em
  \apj}, vol.~158, p.~991, Dec. 1969.

\bibitem{Eling:2005zq}
C.~Eling, ``{Energy in the Einstein-aether theory},'' {\em Phys. Rev.},
  vol.~D73, p.~084026, 2006.
\newblock [Erratum: Phys. Rev.D80,129905(2009)].

\bibitem{Foster:2006az}
B.~Z. Foster, ``{Radiation damping in Einstein-aether theory},'' {\em Phys.
  Rev.}, vol.~D73, p.~104012, 2006.
\newblock [Erratum: Phys. Rev.D75,129904(2007)].

\bibitem{Wald:1993nt}
R.~M. Wald, ``{Black hole entropy is the Noether charge},'' {\em Phys. Rev.},
  vol.~D48, no.~8, pp.~R3427--R3431, 1993.

\bibitem{Iyer:1994ys}
V.~Iyer and R.~M. Wald, ``{Some properties of Noether charge and a proposal for
  dynamical black hole entropy},'' {\em Phys. Rev.}, vol.~D50, pp.~846--864,
  1994.

\bibitem{Carroll:2004st}
S.~M. Carroll, {\em {Spacetime and geometry: An introduction to general
  relativity}}.
\newblock 2004.

\bibitem{Maggiore}
M.~Maggiore, {\em Gravitational Waves: Volume 1: Theory and Experiments}.
\newblock OUP Oxford, 2007.

\bibitem{PW}
{Poisson, E. and Will, C.M.}, {\em Gravity: Newtonian, Post-Newtonian,
  Relativistic}.
\newblock Cambridge University Press, 2014.

\bibitem{Misner:1974qy}
C.~W. Misner, K.~S. Thorne, and J.~A. Wheeler, {\em {Gravitation}}.
\newblock San Francisco: W. H. Freeman, 1973.

\bibitem{Will:1993ns}
C.~M. Will, {\em {Theory and experiment in gravitational physics}}.
\newblock 1993.

\bibitem{Lee:1974pt}
D.~L. Lee, ``{Conservation laws, gravitational waves, and mass losses in the
  Dicke-Brans-Jordan theory of gravity},'' {\em Phys. Rev.}, vol.~D10,
  pp.~2374--2383, 1974.

\bibitem{Lee:1974nq}
D.~L. Lee, A.~P. Lightman, and W.~T. Ni, ``{Conservation laws and variational
  principles in metric theories of gravity},'' {\em Phys. Rev.}, vol.~D10,
  pp.~1685--1700, 1974.

\bibitem{Bekenstein:2004ne}
J.~D. Bekenstein, ``{Relativistic gravitation theory for the MOND paradigm},''
  {\em Phys. Rev.}, vol.~D70, p.~083509, 2004.
\newblock [Erratum: Phys. Rev.D71,069901(2005)].

\bibitem{Moffat:2004nw}
J.~W. Moffat, ``{Modified gravitational theory as an alternative to dark energy
  and dark matter},'' 2004.

\bibitem{Moffat:2007ju}
J.~W. Moffat and V.~T. Toth, ``{Modified Gravity: Cosmology without dark matter
  or Einstein's cosmological constant},'' 2007.

\bibitem{Brill:1964zz}
D.~R. Brill and J.~B. Hartle, ``{Method of the Self-Consistent Field in General
  Relativity and its Application to the Gravitational Geon},'' {\em Phys.
  Rev.}, vol.~135, pp.~B271--B278, 1964.

\bibitem{DeWitt:1960fc}
B.~S. DeWitt and R.~W. Brehme, ``{Radiation damping in a gravitational
  field},'' {\em Annals Phys.}, vol.~9, pp.~220--259, 1960.

\bibitem{MartinGarcia:2008qz}
J.~M. Martin-Garcia, D.~Yllanes, and R.~Portugal, ``{The Invar tensor package:
  Differential invariants of Riemann},'' {\em Comput. Phys. Commun.}, vol.~179,
  pp.~586--590, 2008.

\bibitem{Brizuela:2008ra}
D.~Brizuela, J.~M. Martin-Garcia, and G.~A. Mena~Marugan, ``{xPert: Computer
  algebra for metric perturbation theory},'' {\em Gen. Rel. Grav.}, vol.~41,
  pp.~2415--2431, 2009.

\end{thebibliography}
\bibliographystyle{ieeetr}

\end{document}